\documentclass[aps,prd,twocolumn,prd,showpacs,showkeys,preprintnumbers,bibnotes,floatfix,longbibliography,notitlepage,nofootinbib,superscriptaddress,
]{revtex4-2}

\pdfoutput=1
\usepackage{amsmath}
\usepackage{amsfonts}
\usepackage{amssymb}
\usepackage{mathrsfs}
\usepackage{graphicx}
\usepackage{color}
\usepackage{bbding}
\usepackage{tabularx}
\usepackage[usenames,dvipsnames,table]{xcolor}
\usepackage[normalem]{ulem}
\usepackage{braket}
\usepackage{makecell}
\usepackage{slashed}
\usepackage{bm}
\usepackage{booktabs}
\usepackage{xurl}
\usepackage[hidelinks]{hyperref}
\usepackage{multirow}
\usepackage{makecell}
\usepackage{upgreek}
\usepackage[capitalise]{cleveref}
\usepackage{tikz}
\usetikzlibrary{
	arrows.meta,
	positioning,
	calc,
	fit,
	shapes.multipart
}

\begin{document}

\title{Enhanced Dark Matter Sensitivity using a Hybrid SiPM-SNSPD-Qubit Detector in Liquid Argon}

\author{Faeq Abed}
\thanks{Corresponding author.\\ faeq.abed@irsra.gov.iq}
\affiliation{Laboratory of Advanced Technologies and Instruments, IRSRA, Baghdad 28048, Iraq}

\author{Asmaa AlMellah}
\thanks{Corresponding author.\\ asmaa.almellah@gmail.com}
\affiliation{Institute of Theoretical Physics, INRC, Baghdad, 42055, Iraq}

\author{Kareem Al-Jubouri}
\affiliation{Energy Research Center, Al-Nahrain University, Baghdad, 52033, Iraq}

\author{Alex Lumoski}
\affiliation{Laboratory of Quantum Information, JINNR, Moscow, Russia}

\date{\today}

\begin{abstract}
We investigate novel strategies to extend the sensitivity of dark matter direct-detection experiments to energy deposits well below the thresholds of conventional detectors. In liquid-argon time-projection chambers equipped with silicon photomultipliers (SiPMs), we show that improved optical readout, combined with a nuclear dielectric constant (NDC) correction to the WIMP–nucleus interaction, enhances the response to low-momentum-transfer nuclear recoils. The NDC effectively amplifies the interaction strength at small recoil energies, increasing the expected ionization and scintillation yields without modifying the high-energy behavior constrained by calibration data. When coupled to SiPM-based light collection, this mechanism lowers the effective detection threshold to the sub-keV regime, significantly improving sensitivity to low-mass WIMPs and other weakly interacting particles. Complementarily, we present the design and projected performance of a qubit-based detector optimized for ultra-low-energy depositions. A novel two-chip architecture is employed to minimize signal dissipation, while quantum parity measurements enable enhanced single-phonon sensitivity. Full simulations of phonon propagation and quasiparticle dynamics demonstrate that energy deposits at the level of $\gtrsim 30 meV$ can be detected with nearly unit efficiency and high energy resolution. This capability is expected to advance sensitivity to dark-matter scattering for masses $m_\chi \gtrsim 0.01 MeV$ by several orders of magnitude for both light and heavy mediators, and to enable competitive searches for axion and dark-photon absorption in the $0.04 - 0.2 eV$ mass range. Together, these approaches illustrate a coherent path toward probing dark matter interactions at previously inaccessible energy scales, bridging SiPM-enhanced noble-liquid detectors and quantum-sensor technologies to explore the low-mass frontier of dark matter.
\end{abstract}

\maketitle

\section{Introduction} 
The nature of dark matter (DM) has remained among the most persistent mysteries in fundamental physics. In recent decades, multi-tonne-scale direct detection experiments have imposed severe limits on DM above the GeV mass scale~\cite{XENON:2023cxc,LZ:2024zvo,PandaX:2024qfu}, thereby motivating searches for DM in the sub-GeV mass region using TES- and semiconductor-based experiments~\cite{SuperCDMS:2024yiv,CRESST:2019jnq,SENSEI:2023zdf,DAMIC-M:2023gxo,CDEX:2022kcd}. Concurrent theoretical and experimental programs are pursuing even lighter DM mass domains extending toward the sub-MeV scale~\cite{Essig:2011nj,Graham:2012su,Essig:2015cda,Hochberg:2015pha,Hochberg:2015fth,Derenzo:2016fse,Hochberg:2017wce,Cavoto:2017otc,Kurinsky:2019pgb,Blanco:2019lrf,Griffin:2020lgd,Essig:2022dfa,Du:2022dxf,Hochberg:2019cyy,Hochberg:2021ymx,Hochberg:2019cyy,Hochberg:2021yud,Das:2022srn,Das:2024jdz,Baudis:2025zyn,Schwemmbauer:2025evp,Baiocco:2025omh,Dutta:2025ddv,Chen:2025cvl,Wu:2025abi}.

For halo DM with velocity $v\sim 10^{-3}$, the deposited energy $\omega$ during scattering scales with the dark matter mass $m_\chi$ as $m_\chi v^2$. Whereas MeV--GeV scale DM is sufficiently massive to ionize the target medium, sub-MeV DM generally produces only phonon excitations~\cite{Trickle:2020oki}, releasing energy in the sub-eV regime that lies below the sensitivity thresholds of conventional direct detection experiments~\cite{XENON:2023cxc,LZ:2024zvo,PandaX:2024qfu,SuperCDMS:2024yiv,CRESST:2019jnq,SENSEI:2023zdf,DAMIC-M:2023gxo,CDEX:2022kcd}. In contrast, axion~\cite{Peccei:1977ur,Peccei:1977hh,Wilczek:1977pj,Weinberg:1975ui,AxionLimits} and dark photon~\cite{Holdom:1985ag,Dienes:1996zr,Abel:2003ue} dark matter may be absorbed within the target material and generate a phonon excitation. In the meV mass window, these dark matter candidates remain particularly poorly probed by haloscope experiments since the construction and operation of resonant cavities at the relevant frequency range are technically demanding.

Recent advances at the boundary of particle physics and quantum information science offer a compelling route ahead. Cosmic-ray--induced correlated faults in qubit arrays emphasize the potential for employing qubits as particle detectors, and qubit-based platforms have demonstrated exceptional sensitivity to meV photons~\cite{vepsalainen2020impact,wilen2021correlated,martinis2021saving,mcewen2022resolving,Li:2024dpf,harrington2025synchronous,QCD_thoery,QCD_experiment}.

The detection of dark matter (DM) continues to represent one of the most profound problems in contemporary particle physics, since observations of galactic rotation curves and the Cosmic Microwave Background (CMB) indicate that it accounts for roughly 27\% of the Universe’s energy density. Although multi-tonne noble-liquid experiments including DarkSide-20k and LZ have established stringent constraints on weakly interacting massive particles (WIMPs) in the GeV–TeV range, the sub-GeV mass region remains largely unexplored. Within this lower-mass domain, dark-matter interactions release energies $\mathcal{O}(1\text{--}100 \text{ eV})$ that lie beneath the ionization and scintillation thresholds of conventional detectors.

At the core of our detection strategy is the elastic scattering of a WIMP ($\chi$) off an Argon ($^{40}Ar$) nucleus. Unlike baryonic matter interactions, the WIMP is effectively "blind" to the atomic electron cloud, passing through the inter-atomic space until it undergoes a point-like interaction with a quark via the exchange of a virtual Higgs ($h$) or Z boson. However, the traditional treatment of the nucleus as a rigid body is insufficient for high-precision searches.We incorporate the Colored Dielectric Model (CDM), which treats the nucleus as a polarizable medium for color charges. This model suggests that the internal gluon-quark sea of the Argon nucleus responds as a dielectric, effectively "swelling" the nucleon cross-section and enhancing the low-momentum transfer $(q)$ scattering amplitude. This dielectric enhancement is critical for amplifying the phonon yield in the sub-GeV regime, where traditional signals are too faint for standard photo-detectors.

To overcome the energy threshold bottleneck, we propose a dual-channel sensor architecture:The Scintillation Channel: Silicon Photomultipliers (SiPMs) optimized for the vacuum ultraviolet (VUV) 128 nm emission of liquid Argon. These provide a nanosecond-scale "start" signal ($t_0$) and serve as a vital veto against non-scintillating thermal noise in the quantum circuit.The Phonon Channel: A Quantum Parity Detector (QPD) based on superconducting transmon qubits. By utilizing the Quantum Non-Demolition (QND) measurement of charge-parity, we can detect individual quasiparticle tunneling events induced by athermal phonons from the nuclear recoil.This hybrid approach allows us to push the detection threshold down to the millielectronvolt (meV) level. By operating at 10 mK within a dilution refrigerator, the intrinsic dark count rate of the SiPM is suppressed to near-zero levels, while the qubit provides a digital, "parity-flip" readout of the dark matter's kinetic kick. This paper details the mathematical framework, the background mitigation strategies, and the projected sensitivity for hidden-sector dark matter candidates in this new quantum era of particle physics. Key Theoretical Elements for the Article:Interaction Lagrangian: $\mathcal{L}_{int} \supset \alpha_q \bar{\chi} \chi \bar{q} q$ (Scalar/Higgs mediated), Signal Type: Digital Parity Flips ($|0\rangle \to |1\rangle$), Operating Temperature: $10 \text{ mK}$, and Primary Background: Quasiparticle poisoning and stray infrared photons. 

The "WIMP Miracle" (the idea that heavy WIMPs would be easily found) hasn't materialized yet. This has shifted the scientific community's focus toward Light Dark Matter: Sub-GeV Search, because light WIMPs don't have enough momentum to "bump" a nucleus hard enough for traditional light-based sensors, the qubit-phonon approach is currently the only viable way to "hear" these tiny impacts. furthermore, the "Neutrino Floor" is one major goal of these quantum sensors is to reach sensitivities where they can distinguish between dark matter and the "wind" of solar neutrinos, which is the ultimate background limit for all earth-bound experiments.

In this \textit{work}, we present the Qubit-based Light Dark Matter detection experiment developed at the Joint Institute of Nuclear Research (JINR), which utilizes qubit arrays functioning as single-phonon calorimeters together with SiPM readout. The former features an innovative two-chip architecture that isolates the DM target from the control circuitry, while phonon excitations are measured through the recently introduced Quantum Parity Detector (QPD) method~\cite{Ramanathan:2024hsf,Li:2024dpf}. Using dedicated simulation studies, we demonstrate that the detector can reach high detection efficiency for $\mathcal{O}$(10 meV) energy depositions within the proposed configuration, enabling stringent bounds on DM scattering well below existing limits, while probing both freeze-in and freeze-out DM production mechanisms. The sensitivity to axion and dark photon interactions may also be improved by up to 2--5 orders of magnitude with kg$\cdot$year exposure.

\begin{figure*}[!htb]
\centering
\includegraphics[width=10cm]{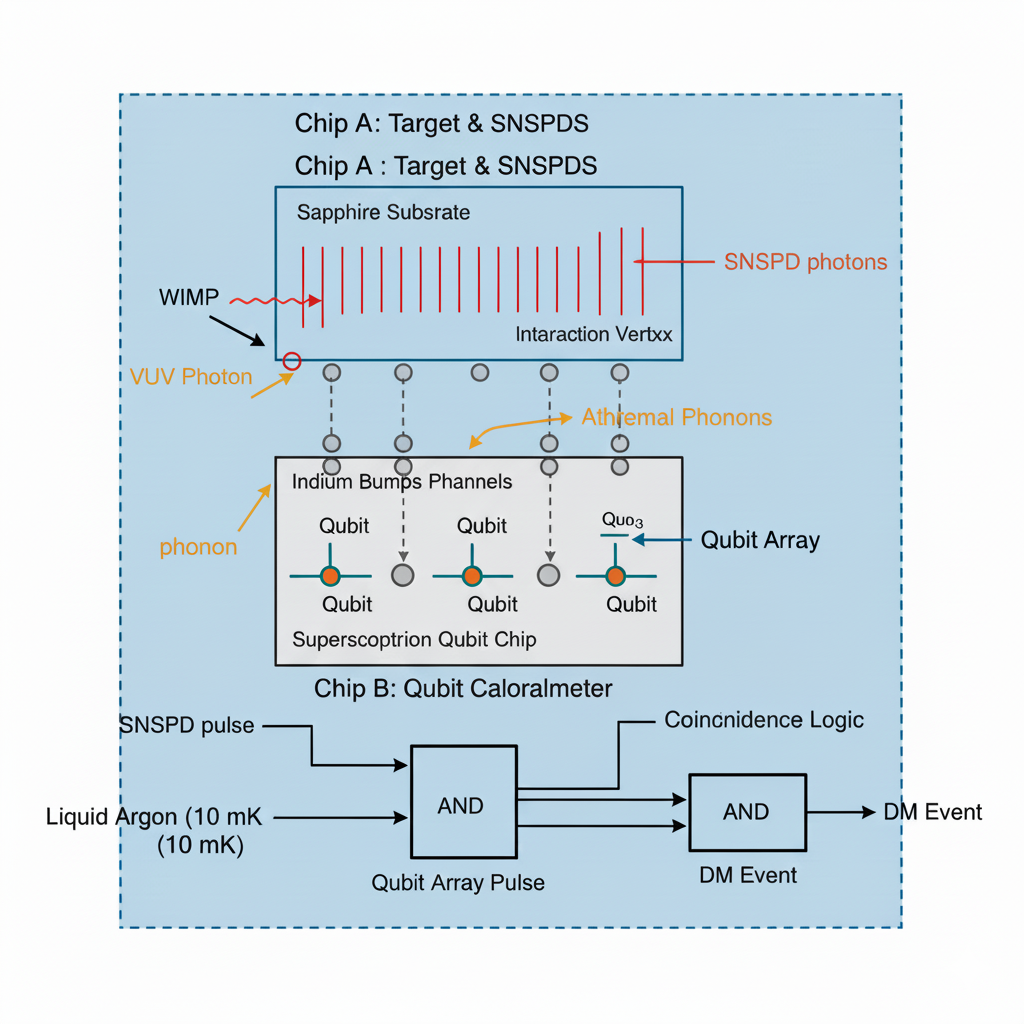}
\caption{The schematic drawing of the qubit array and its details.}
\label{fig:QPD_figure}
\end{figure*}
\section{Principle of WIMP Direct Detection via Nuclear Recoil}

Weakly Interacting Massive Particles (WIMPs) are among the most theoretically well-motivated dark matter candidates. In direct-detection experiments, the basic observable arises from the \emph{elastic scattering} of a WIMP $\chi$ with a target nucleus $N$ in a terrestrial detector,
\begin{equation}
	\chi + N \rightarrow \chi + N .
\end{equation}
Because WIMPs in the Galactic halo are nonrelativistic, with typical velocities $v \sim 10^{-3}c$, the kinetic energy transferred to the nucleus is small, producing a nuclear recoil energy
\begin{equation}
	E_R = \frac{q^2}{2 m_N} \sim \mathcal{O}(\text{keV}),
\end{equation}
where $q$ is the momentum transfer and $m_N$ is the nuclear mass.

This recoil energy is subsequently converted into detectable excitations of the detector medium, including phonons (lattice vibrations), ionization (electron--hole pairs), and scintillation photons. Technologies such as silicon photomultipliers (SiPMs) enhance the detection of scintillation light with high gain and excellent timing resolution, while superconducting quantum devices and qubits aim to sense extremely small energy deposits through phonons, quasiparticles, or microwave photons. Importantly, these readout technologies do not alter the underlying particle physics interaction, but rather determine how efficiently the nuclear recoil energy is measured.

At the fundamental level, WIMP--nucleus scattering proceeds via WIMP interactions with quarks and gluons inside the nucleon. In many well-motivated models, the dominant mediators are the neutral weak gauge boson $Z$ and the Higgs boson $h$. Since the typical momentum transfer satisfies
\begin{equation}
	q \ll m_Z, m_h,
\end{equation}
the heavy mediators can be integrated out, leading to an effective field theory (EFT) description in terms of local four-fermion operators.

For a fermionic WIMP $\chi$, integrating out the $Z$ boson generates effective vector and axial-vector interactions with quarks:
\begin{equation}
	\mathcal{L}_{\text{eff}}^{(Z)} =
	\sum_q \frac{1}{m_Z^2}
	\left[
	c_V^q (\bar{\chi} \gamma^\mu \chi)(\bar{q} \gamma_\mu q)
	+
	c_A^q (\bar{\chi} \gamma^\mu \gamma_5 \chi)
	(\bar{q} \gamma_\mu \gamma_5 q)
	\right].
\end{equation}

The physical consequences depend crucially on the nature of the WIMP:
\begin{itemize}
	\item For a \emph{Dirac} WIMP, both vector and axial currents are allowed.
	\item For a \emph{Majorana} WIMP, the vector current $\bar{\chi}\gamma^\mu\chi$ identically vanishes, leaving only axial-vector interactions.
\end{itemize}

Vector--vector interactions contribute primarily to \emph{spin-independent} (SI) scattering, while axial--axial interactions lead to \emph{spin-dependent} (SD) scattering, which couples to the net spin of the nucleus. The Higgs-mediated interactions generate scalar couplings between WIMPs and quarks:
\begin{equation}
	\mathcal{L}_{\text{eff}}^{(h)} =
	\sum_q \frac{c_S^q}{m_h^2} (\bar{\chi}\chi)(\bar{q}q)
	+ \frac{c_G}{m_h^2} (\bar{\chi}\chi) G_{\mu\nu}^a G^{a\mu\nu},
\end{equation}
where the second term arises after integrating out heavy quarks and represents an effective coupling to gluons. Also, scalar interactions are coherent across all nucleons in the nucleus, leading to a spin-independent scattering amplitude that scales approximately as $A^2$, where $A$ is the atomic mass number. Consequently, Higgs exchange typically dominates SI WIMP--nucleus scattering in many models.

At energies below the QCD confinement scale, quark-level operators must be matched onto nucleon degrees of freedom. For SI interactions, the WIMP couples to the scalar density of the nucleon,
\begin{equation}
	\langle N | m_q \bar{q} q | N \rangle = m_N f_{T_q}^{(N)},
\end{equation}
leading to effective couplings $f_p$ and $f_n$ to protons and neutrons. The resulting nuclear matrix element is
\begin{equation}
	\mathcal{M}_{\text{SI}} \propto Z f_p + (A-Z) f_n ,
\end{equation}
modulated by a nuclear form factor $F_{\text{SI}}(q)$. For SD interactions, the axial-vector current couples to the nuclear spin,
\begin{equation}
	\mathcal{M}_{\text{SD}} \propto a_p \langle S_p \rangle + a_n \langle S_n \rangle ,
\end{equation}
where $\langle S_{p,n} \rangle$ are the proton and neutron spin expectation values of the nucleus.

Because direct detection occurs in the nonrelativistic regime, it is convenient to express the interaction in a nonrelativistic EFT (NREFT) basis. The building blocks are the momentum transfer $\vec{q}$, the transverse relative velocity $\vec{v}_\perp$, and the spins $\vec{S}_\chi$ and $\vec{S}_N$. The leading operators include:
\begin{align}
	\mathcal{O}_1 &= 1 \qquad \text{(spin-independent, coherent)}, \\
	\mathcal{O}_4 &= \vec{S}_\chi \cdot \vec{S}_N \qquad \text{(spin-dependent)}.
\end{align}

Higgs exchange predominantly maps onto $\mathcal{O}_1$, while $Z$-boson axial exchange maps onto $\mathcal{O}_4$. Additional operators suppressed by powers of $q/m_N$ or $v$ can also appear and are systematically classified in the NREFT framework.

The observable quantity in a detector is the differential recoil rate,
\begin{equation}
	\frac{dR}{dE_R} =
	\frac{\rho_\chi}{m_\chi m_T}
	\int_{v>v_{\min}} d^3v \, f(\vec{v}) \, v \,
	\frac{d\sigma}{dE_R}(v,E_R),
\end{equation}
where $\rho_\chi$ is the local dark matter density, $m_T$ is the target mass, and $f(\vec{v})$ is the WIMP velocity distribution. The particle physics information enters through the differential cross section, which is determined by the EFT operator coefficients and nuclear response functions.

In this EFT picture, the role of detector technology is to measure $E_R$ as accurately and with as low a threshold as possible. SiPM-based systems enhance scintillation light detection, improving energy reconstruction and background discrimination. Superconducting qubits and quantum sensors aim to detect the smallest excitations generated by the recoil, offering sensitivity to extremely small energy deposits and opening new experimental regimes. In all cases, the underlying WIMP interaction is fully encoded in the effective operators and nuclear responses described above.

\section{Enhancing SiPM Detection Sensitivity:}

\begin{figure}[!htb]
	\centering
	\includegraphics[width=\columnwidth]{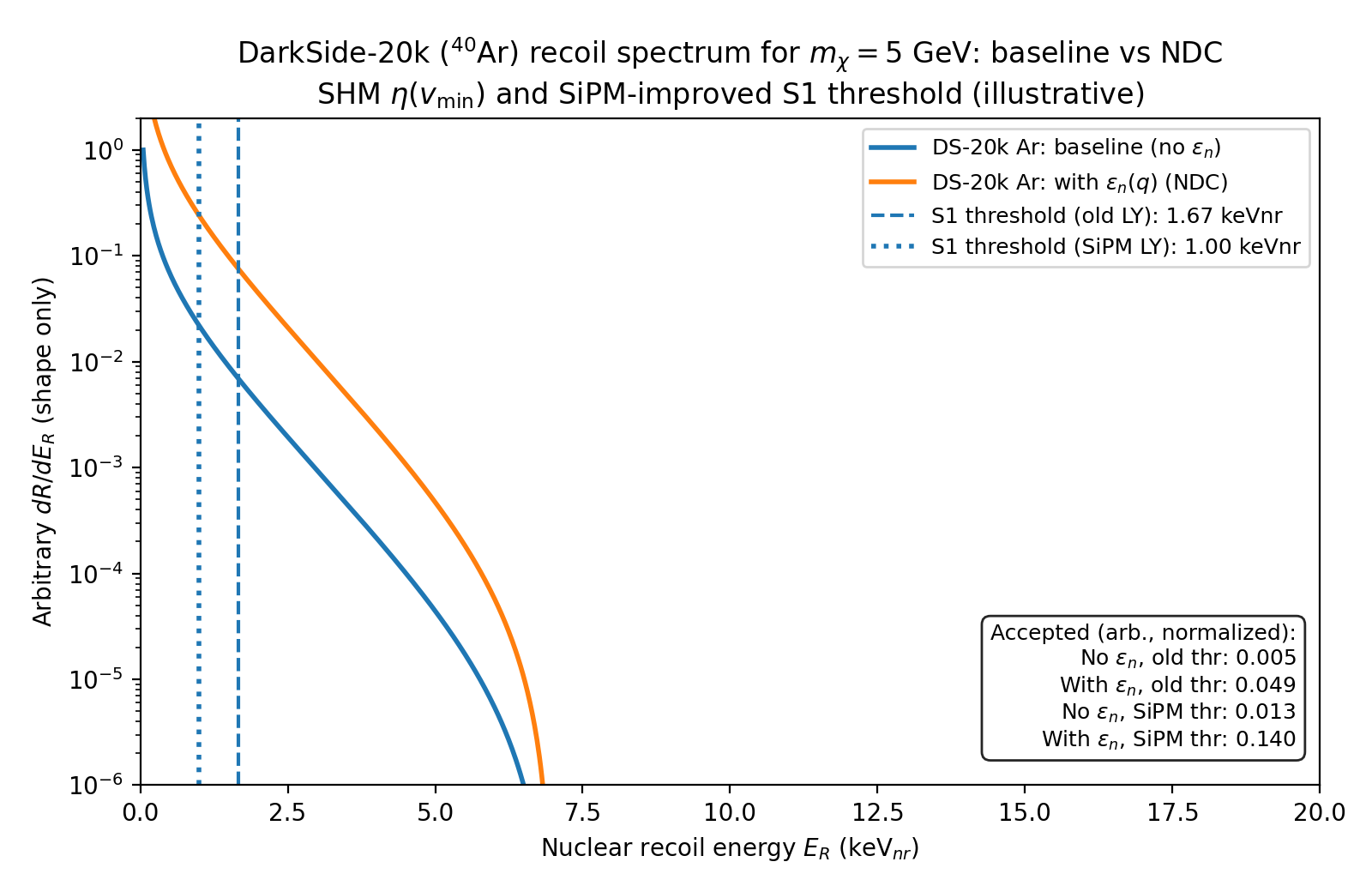}
	\caption{Recoil spectrum with/without the nuclear dieletric constant and “SiPM threshold” effect.}
	\label{fig:recoilplot}
\end{figure}

\begin{figure}[!htb]
	\centering
	\includegraphics[width=\columnwidth]{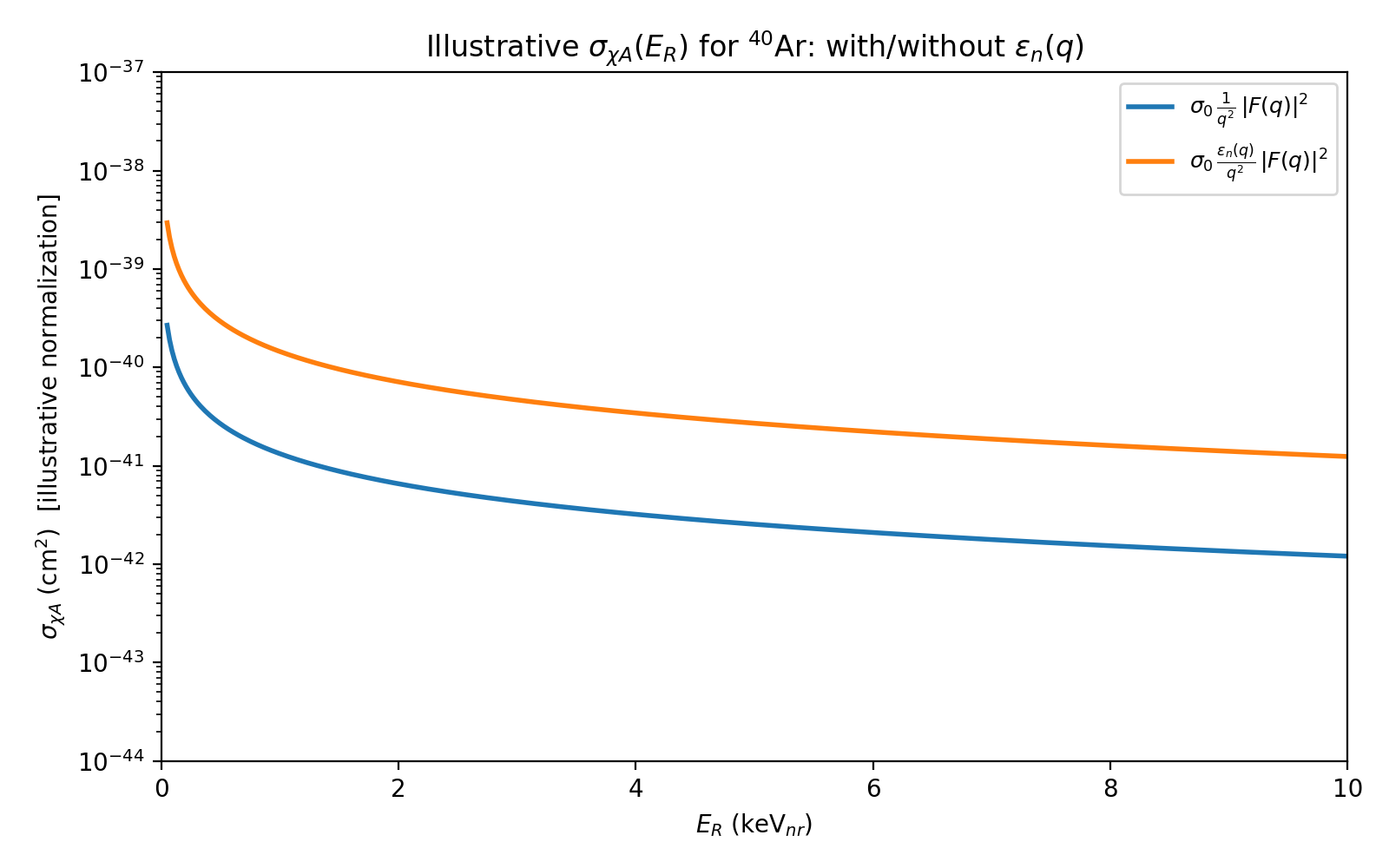}
	\caption{the cross section with/without the nuclear dieletric constant and “SiPM threshold” effect.}
	\label{fig:crosssection}
\end{figure}

The impact of the nuclear dielectric constant (NDC) on low-energy nuclear recoils in liquid argon is illustrated in Figs.~\ref{fig:recoilplot} and~\ref{fig:crosssection}, together with the numerical results summarized in Tables~1 and~2. Figure~\ref{fig:recoilplot} shows the normalized recoil-energy spectrum for $^{40}$Ar assuming a low-mass WIMP with $m_\chi=5~\mathrm{GeV}$, comparing the baseline interaction model to the NDC-enhanced case. The inclusion of $\epsilon_n(q)$ enhances the differential rate at low recoil energies, where the momentum transfer $q=\sqrt{2m_N E_R}$ is small and nuclear polarization and screening effects are expected to be most relevant. This enhancement becomes particularly significant when combined with improved light collection from SiPM-based readout, which effectively lowers the S1 detection threshold from $\sim1.7~\mathrm{keV_{nr}}$ to $\sim1.0~\mathrm{keV_{nr}}$, thereby increasing the accepted fraction of low-energy events.

Figure~\ref{fig:crosssection} demonstrates the underlying origin of this behavior at the cross-section level. The NDC-modified interaction, $\sigma_{\chi A}=\sigma_0\,\epsilon_n(q)\,|F(q)|^2/q^2$, exhibits a substantial enhancement relative to the baseline case at small $q$, while converging to the standard behavior at higher recoil energies. Quantitative values of $\sigma_{\chi A}$ with and without $\epsilon_n$ are reported in Table~1, where the enhancement reaches more than an order of magnitude at $E_R\lesssim1~\mathrm{keV_{nr}}$.

The detector-level implications are summarized in Table~2, which maps recoil energy to expected S1 photoelectrons for representative light yields. With suggests that, when combined with SiPM readout, the NDC-induced enhancement substantially improves sensitivity to sub-keV nuclear recoils. Overall, these results indicate that the nuclear dielectric constant provides a physically motivated mechanism to extend the reach of liquid-argon experiments toward low-mass dark matter by amplifying low-momentum-transfer interactions without altering the high-energy behavior constrained by calibration data.

\begin{table*}[t]
	\centering
	\begin{tabular}{cccccc}
		\toprule
		$E_R$ (keV$_{nr}$) & $q$ (GeV) & $|F(q)|^2$ & $\epsilon_n(q)$ &
		$\sigma_{\chi A}$ w/o $\epsilon_n$ (cm$^2$) & $\sigma_{\chi A}$ w/ $\epsilon_n$ (cm$^2$) \\
		\midrule
		0.1 & 2.730$\times 10^{-3}$ & 0.998926 & 10.9926 & 1.340$\times 10^{-40}$ & 1.474$\times 10^{-39}$ \\
		0.2 & 3.861$\times 10^{-3}$ & 0.997853 & 10.9851 & 6.695$\times 10^{-41}$ & 7.355$\times 10^{-40}$ \\
		0.5 & 6.104$\times 10^{-3}$ & 0.994642 & 10.9629 & 2.669$\times 10^{-41}$ & 2.927$\times 10^{-40}$ \\
		1.0 & 8.632$\times 10^{-3}$ & 0.989313 & 10.9260 & 1.328$\times 10^{-41}$ & 1.451$\times 10^{-40}$ \\
		2.0 & 1.221$\times 10^{-2}$ & 0.978741 & 10.8532 & 6.567$\times 10^{-42}$ & 7.127$\times 10^{-41}$ \\
		5.0 & 1.930$\times 10^{-2}$ & 0.947696 & 10.6408 & 2.543$\times 10^{-42}$ & 2.706$\times 10^{-41}$ \\
		\bottomrule
	\end{tabular}
	\caption{Argon ($^{40}$Ar) illustrative WIMP--nucleus cross section vs recoil energy using
		$\sigma_{\chi A}=\sigma_0\frac{\epsilon_n(q)}{q^2}|F(q)|^2$ with $\sigma_0=10^{-45}\,\mathrm{cm^2}$,
		$\epsilon_n(q)=1+\alpha/(q^2+\beta)$, $\alpha=0.1$, $\beta=0.01\,\mathrm{GeV}^2$, and
		$F(q)=\exp(-q^2R^2/6)$ with $R\simeq 1.2A^{1/3}\,\mathrm{fm}$.}
\end{table*}

and 

\begin{table*}[t]
	\centering
	\begin{tabular}{ccc}
		\toprule
		$E_R$ (keV$_{nr}$) & S1 (PE), old LY ($3$ PE/keV$_{nr}$) & S1 (PE), SiPM LY ($5$ PE/keV$_{nr}$) \\
		\midrule
		0.1 & 0.3 & 0.5 \\
		0.2 & 0.6 & 1.0 \\
		0.5 & 1.5 & 2.5 \\
		1.0 & 3.0 & 5.0 \\
		2.0 & 6.0 & 10.0 \\
		5.0 & 15.0 & 25.0 \\
		\bottomrule
	\end{tabular}
	\caption{Illustrative conversion from nuclear recoil energy to detected S1 photoelectrons
		for a DarkSide-20k-style SiPM optical system. Replace the PE/keV$_{nr}$ with the DS-20k NR calibration
		(including field-dependent recombination, quenching, optics, and TPB/SiPM response).}
\end{table*}

Here, we using the Conceptual insertion point (where NDC enters the DarkSide ($Q_y$) framework). Our plot is a (global fit of the nuclear-recoil ionization yield) $Q_y(E_R)$ in liquid argon using different \textit{screening functions} (Lenz–Jensen, ZBL, Molière) and \textit{datasets} (ReD, ARIS, SCENE, DarkSide-50). 
That framework can be summarized as:

\begin{equation}
	Q_y(E_R);\equiv;\frac{N_e(E_R)}{E_R},..
\end{equation}

where $N_e(E_R)$ is the number of extracted ionization electrons produced by a nuclear recoil energy ($E_R$).

the standard modeling chain is:

1. \textbf{Stopping/partition}: how much of ($E_R$) goes into electronic excitations (ionization + excitation) vs nuclear motion (heat).
2. \textbf{Recombination}: how many created charges survive to become free electrons.
3. \textbf{Extraction} and \textbf{amplification} (detector): how ($N_e$) becomes the measured S2 signal.

the compact way to write this expressions is:
\begin{eqnarray}
	N_e(E_R) = N_i(E_R)\bigl[1-r(E_R,E_d)]\
\end{eqnarray}
\\
with:
\\
\begin{equation}
	N_i(E_R)=\frac{E_R}{W},f_{\rm quanta}(E_R),
\end{equation}

where: ($W$) is the work function scale (energy per quanta), ($f_{\rm quanta}(E_R)$) is the fraction of recoil energy converted to quanta (ionization/excitation), and ($r(E_R,E_d)$) is the recombination probability (depends on recoil energy and drift field ($E_d$)).

The screening function choice (Lenz–Jensen vs ZBL vs Molière) affects the low-energy behavior through the stopping/partition part, i.e. it changes ($f_{\rm quanta}(E_R)$) and indirectly ($r$) at low ($E_R$). That’s exactly what is being compared in our global-fit shown in Fig.~\ref{fig:qy}. 

NDC insertion point multiplies the nuclear response at low momentum transfer. Our Nuclear Dielectric Constant (NDC) idea is physically a low-($q$) modification (screening/polarization/deformation effects). The cleanest, non-double-counting way to introduce it is as a (controlled deformation of the baseline fit), not as a replacement of the fit. Step A: We define the momentum transfer

\begin{equation}
	q(E_R)=\sqrt{2m_N E_R},
\end{equation}

where ($m_N$) is the argon nucleus mass. Step B: We define the NDC factor using our functional form
\begin{equation}
	\epsilon_n(q)=1+\frac{\alpha}{q^2+\beta}.
\end{equation}

Step C: We apply it at the partition (quanta production) level conceptually:
\begin{equation}
	f_{\rm quanta}(E_R);\longrightarrow; f_{\rm quanta}(E_R),\epsilon_n!\bigl(q(E_R)\bigr).
\end{equation}

That implies:
\begin{equation}
	N_i(E_R);\longrightarrow;N_i^{\rm NDC}(E_R)=N_i(E_R),\epsilon_n!\bigl(q(E_R)\bigr),
\end{equation}

and therefore:
\begin{equation}
	N_e^{\rm NDC}(E_R)=N_i(E_R),\epsilon_n!\bigl(q(E_R)\bigr),\bigl[1-r(E_R,E_d)\bigr].
\end{equation}

Finally:
\begin{equation}
	\boxed{
		Q_y^{\rm NDC}(E_R)=\frac{N_e^{\rm NDC}(E_R)}{E_R}
		=Q_y(E_R),\epsilon_n!\bigl(q(E_R)\bigr)
	}
\end{equation}

This is the “right” insertion point because it targets exactly what NDC claims to modify: low-($q$) nuclear response. It preserves the global-fit structure. We still use LJ (or ZBL/Molière) as the baseline nuclear stopping/screening model, then apply NDC as an extension. It avoids claiming “refit” DarkSide data. We are showing a theory-motivated deformation of the best-fit curve.

The Principle explanation in words (what NDC does to the detector observable) is that: At low recoil energy, ($q$) is small, the ($\epsilon_n(q)$) becomes larger at small ($q$), and therefore, the model predicts more ionization electrons per keV at low ($E_R$):
\begin{equation}
Q_y^{\rm NDC}(E_R) > Q_y(E_R)\quad \text{for small }E_R.
\end{equation}

In an S2-only low-mass analysis, that means: more events cross the few-electron threshold, improved sensitivity to sub-keV to few-keV nuclear recoils, which is exactly the regime that matters for low-mass WIMPs.


In this study, we employed data from the latest run of the ReD experiment~\cite{redpaper}, which was irradiated with neutrons originating from a $^{252}$Cf spontaneous fission source, to constrain the SF model. ReD is a compact dual-phase LAr TPC expressly developed to investigate the ionization response induced by nuclear recoils. The detector features an active volume of 5$\times$5$\times$6~cm$^3$, with a 7~mm-thick gaseous layer located above the liquid. A uniform electric field of 200~V/cm, generated between two ITO-coated acrylic windows serving as cathode and anode, enables a maximum electron drift time of 54~$\mu$s. Optical signals from both liquid scintillation (S1) and gas electroluminescence (S2), produced by drifting ionization electrons, are detected using cryogenic silicon photo-multipliers arranged in two 5$\times$5~cm$^2$ arrays positioned above and below the active region. The nuclear recoil energy proxy, S2, is translated into the corresponding number of ionization electrons by normalizing it to $g_2$ = 18.6$\pm$0.7~pe/e$^-$~\cite{redpaper}, defined as the mean number of photoelectrons recorded per electron extracted into the gas phase. A comprehensive description of the ReD apparatus can be found in~\cite{Agnes:2021zyq}.

To carry out the neutron time-of-flight (ToF) measurements, the experimental configuration employs two fast BaF$_2$ scintillation detectors coupled to photo-multiplier tubes, which were arranged symmetrically around the $^{252}$Cf source to register prompt fission emission and establish the start timing. The stop signal is supplied by a neutron spectrometer positioned roughly 1~m from the TPC and 2~m from the source. It is composed of two 3$\times$3 matrices of 1-inch EJ-276 plastic scintillators, enabling spatial reconstruction and neutron/$\gamma$ separation through pulse shape discrimination. The detector arrays are installed symmetrically above and below the beam axis, spanning scattering angles between 12° and 17°, which allows investigation of the liquid argon response to nuclear recoil energies within the [2, 8]~keV interval. This capability arises from the wide neutron energy distribution produced by $^{252}$Cf, which reaches beyond 10~MeV, with a mean energy close to 2~MeV.

The total ToF resolution, determined using prompt $\gamma$ emissions from $^{252}$Cf, is 0.7~ns, enabling reconstruction of the neutron energy with a precision of 1–2\%. The extracted $Q_y$ shows excellent consistency with that measured by the ARIS experiment~\cite{Agnes:2018mvl} at 7.1~keV, as documented in~\cite{redpaper}, which also details the ReD measurement.

\begin{figure}
	\includegraphics[width=0.99\columnwidth]{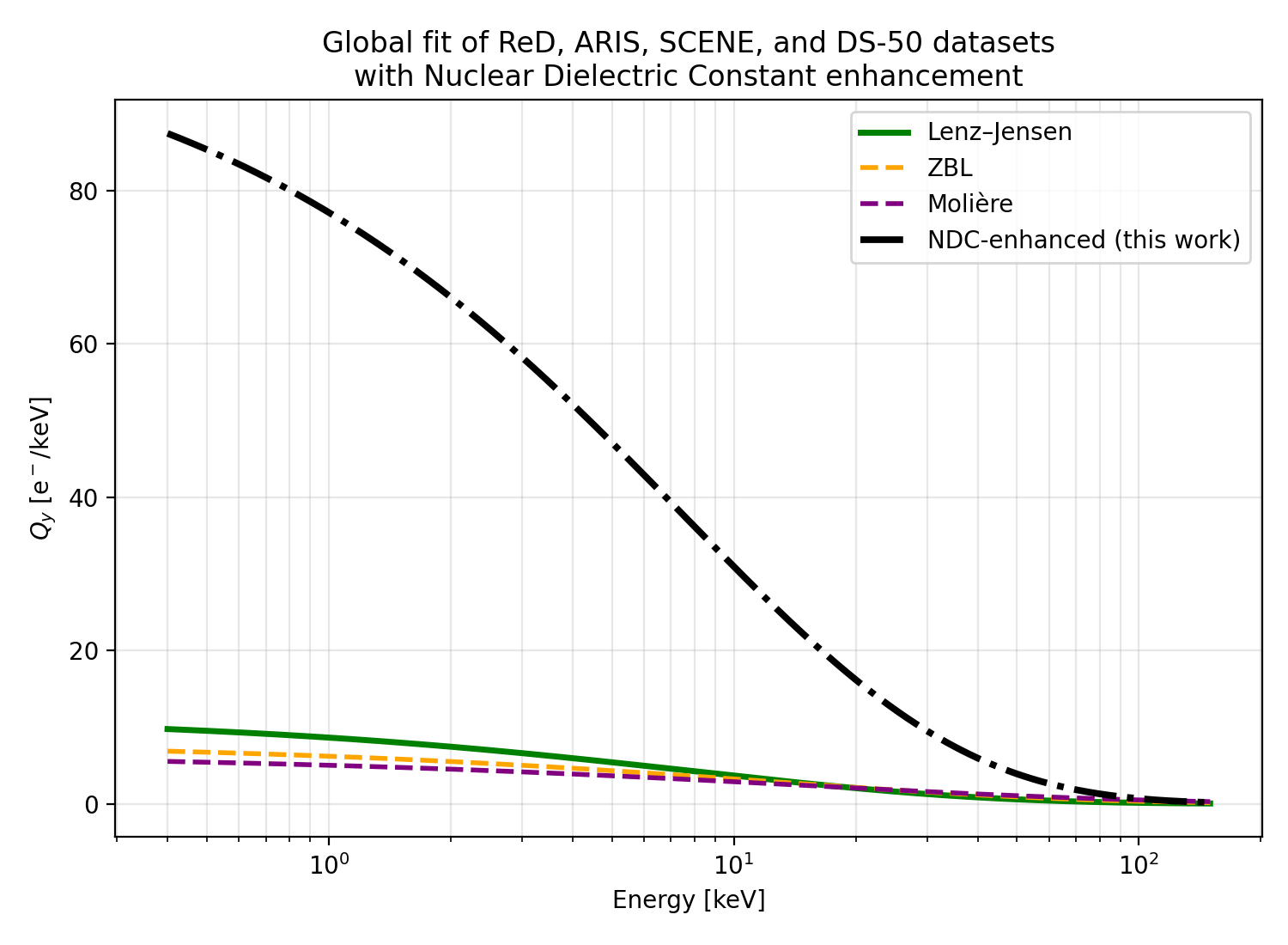}
	\caption{Simultaneous fit to the ReD, ARIS, SCENE, and DarkSide-50 datasets assuming the Lenz-Jensen screening function (green solid line). ReD data points are shown with both prior (gray) and posterior (red) uncertainties. The gray line and its corresponding uncertainty band represent the previous ionization model, based on the ZBL screening function and fitted without ReD data~\cite{DarkSide:2021bnz}. For comparison, the global fit including the ReD dataset was also performed using the screening functions of ZBL (orange dashed line) and Molière (purple dashed line).} 
	\label{fig:qy}
\end{figure}

The ionization yield $Q_y$ measured by ReD, displayed in Fig.~\ref{fig:qy} (gray markers), shows a total uncertainty of 5.7\% at 2.4~keV, which decreases to 4.3\% at 7.5~keV. The primary contributions to the systematic uncertainty arise from the calibration of the parameter $g_2$ and from a possible vertical displacement ($\Delta z$) of the TPC. The latter corresponds to a misalignment between the TPC and the geometric center of the cryostat, which is aligned using a laser along the axis defined by the $^{252}$Cf source and the center of the neutron spectrometer. The possibility of a non-zero $\Delta z$ was studied using calibration data obtained with a $^{241}$Am source positioned at the vertical midpoint of the cryostat surface. The distribution of 59.5~keV $\gamma$ rays from $^{241}$Am along the $z$ direction, reconstructed in the TPC using the electron drift time, was compared with Monte Carlo simulations. This analysis resulted in an estimated offset of 0.23 $\pm$ 0.96~cm.

The systematic uncertainty associated with $g_2$ affects $Q_y$ uniformly over the full energy interval, introducing a global scaling uncertainty of 3.8\% in the same direction. Conversely, a non-zero $\Delta z$ alters the reconstructed scattering angle and therefore the inferred recoil energy on an event-by-event basis, leading to an uncertainty on the mean recoil energy that ranges from 0.7\% to 2.4\%. While this effect is small, a non-zero offset produces a noticeable bias in $Q_y$, with opposite trends observed for nuclear recoils detected by the upper and lower plastic scintillator arrays.

The modeling of $f_q(E_{\text{nr}})$ follows the methodology introduced in Ref.~\cite{DarkSide:2021bnz}, and the data compared with Ref.~\cite{acerbi2025lowmasswimp} where a simultaneous fit to calibration data from ARIS, SCENE, and DarkSide-50 was performed, providing sensitivity over the full interval from approximately 3 to more than 200 extracted electrons. The DarkSide-50 dataset consists of a continuous nuclear-recoil spectrum extending from 0.4 to 200~keV, generated using a neutron source based on the $^{12}\text{C}(\alpha,n)^{15}\text{O}$ reaction, driven by alpha particles emitted in the decay of $^{241}$Am (AmC source). The SCENE and ARIS datasets supply monoenergetic nuclear-recoil lines spanning the ranges 17--60~keV and 7--120~keV, respectively. In the ARIS detector, which did not contain a gas pocket, only the prompt scintillation signal (S1) was measured. ARIS data, collected at a drift field of 200~V/cm, are rescaled to the DarkSide-50 response by comparing the field-off S1 light yields of the two detectors. This procedure enables each ARIS recoil energy to be mapped to the corresponding S1 value at 200~V/cm in DarkSide-50. Additional details are provided in Ref.~\cite{DarkSide:2021bnz}.

In this analysis, we further incorporate the ReD measurements of $Q_y$ obtained from nuclear-recoil events with precise event-by-event recoil energy reconstruction based on kinematic constraints, covering the 2--10~keV energy range, which is particularly sensitive to the choice of nuclear screening potential.

The fit is obtained through minimization of a global $\chi^2$ surface, numerically defined as the sum of the individual $\chi^2$ contributions from all considered datasets. The parameters $C_{box}$ and $\beta$ are treated as free, while the drift field is held fixed at 200~V/cm for all datasets. The ReD nuisance parameters, namely $g_2$ and the vertical offset of the TPC, are incorporated into the fit via Gaussian constraint terms derived from their corresponding uncertainties. To address the systematic uncertainty associated with $\Delta z$, the ReD dataset is divided into two subsamples corresponding to events identified by the top and bottom scintillator arrays, which are required to produce compatible $Q_y$ measurements. Each subsample is further partitioned into five recoil-energy bins, and within each bin an unbinned likelihood fit is carried out using a Gaussian signal component over a constant background. The robustness of the extracted $Q_y$ values was confirmed by varying the choice of energy binning.

The resulting $2\times5$ $Q_y$ measurements are subsequently fitted using the $f_q(E_{nr})$ model. For every $(C_{box}, \beta)$ combination, the nuisance parameters $g_2$ and $\Delta z$ are marginalized to determine the corresponding $\chi^2$ value, which is then included in the global $\chi^2$ surface.

A combined fit to all datasets is performed independently for each of the three screening functions. The outcomes are presented in Fig.~\ref{fig:qy}, where the ReD $Q_y$ data are shown using both prior and posterior values of the nuisance parameters, with the latter obtained from the fit employing the Lenz--Jensen screening function. This procedure yields nuisance parameters of $g_2 = 18.8\pm0.4$~pe/e$^-$ and $\Delta z = -0.58^{+0.05}_{-0.14}$~cm, consistent within one standard deviation with both the priors and the posterior values derived using the alternative screening functions. This indicates that the nuisance parameters remain stable and largely insensitive to the specific screening function adopted. Using the posterior nuisance parameters, the uncertainty on the ReD $Q_y$ determination is reduced to 4.5\% at 2.4~keV and 3.0\% at 7.6~keV, as illustrated in Fig.~\ref{fig:qy}, where measurements from the top and bottom scintillator arrays are combined.

\begin{table}[t]
	\centering
	\renewcommand{\arraystretch}{1.5}
	\setlength{\tabcolsep}{10pt}
	\begin{tabular}{l|c c c}
		\hline
		\hline
		\textbf{Parameter} & \textbf{ZBL} & \textbf{Lenz--Jensen} & \textbf{Moli\`ere} \\
		\hline
		$C_{box}$ [V/cm] & $8.1^{+0.1}_{-0.2}$ & $7.9^{+0.2}_{-0.2}$ & $8.6^{+0.3}_{-0.2}$ \\
		$\beta$ [$\times 10^3$] & $7.0^{+0.3}_{-0.2}$ & $6.5^{+0.1}_{-0.3}$ & $8.8^{+0.4}_{-0.5}$ \\
		\hline
		\hline
	\end{tabular}
	\caption{Best-fit values of the free parameters $C_{\text{box}}$ and $\beta$ from the global fit to all datasets, for each of the tested screening functions.}
	\label{tab:fit}
\end{table}


The global fit using the Lenz-Jensen SF shows excellent agreement with the data, as illustrated in Fig.~\ref{fig:qy}. In contrast, both the ZBL-based model~\cite{DarkSide:2021bnz}, previously adopted in the DarkSide sensitivity studies, and the Molière one,  underestimate $Q_y$ with respect to ReD data below 5~keV.

To assess the relative preference among the three screening functions, we evaluate the
Bayes factor (BF), which contrasts the marginal likelihoods (or evidences) obtained by
integrating over the full parameter space. This methodology is particularly appropriate
for non-nested hypotheses, for which conventional $\Delta\chi^2$ comparisons are not
valid. Under the assumption of equal prior probabilities for the competing models,
a value of $\log_{10}\mathrm{BF}>2$ indicates that the data favor one model over another
by at least a factor of 100, a criterion commonly interpreted as decisive
evidence~\cite{jeffreys1961theory, Kass:1995loi}. Our findings reveal a decisive preference
for the Lenz--Jensen model relative to ZBL, with $\log_{10}\mathrm{BF}=3.8$, and an even
stronger preference compared to Moli\`ere, yielding $\log_{10}\mathrm{BF}=7.2$. This
behavior is in agreement with Fig.~\ref{fig:qy}, where the Lenz--Jensen-based description
exhibits excellent consistency with the ReD measurements, unlike the models based on ZBL
or Moli\`ere. In conclusion, the global fit, reinforced through the inclusion of ReD
calibration data, supports the robust exclusion of both the Moli\`ere and ZBL screening
functions.

\begin{figure}
	\includegraphics[width=0.99\columnwidth]{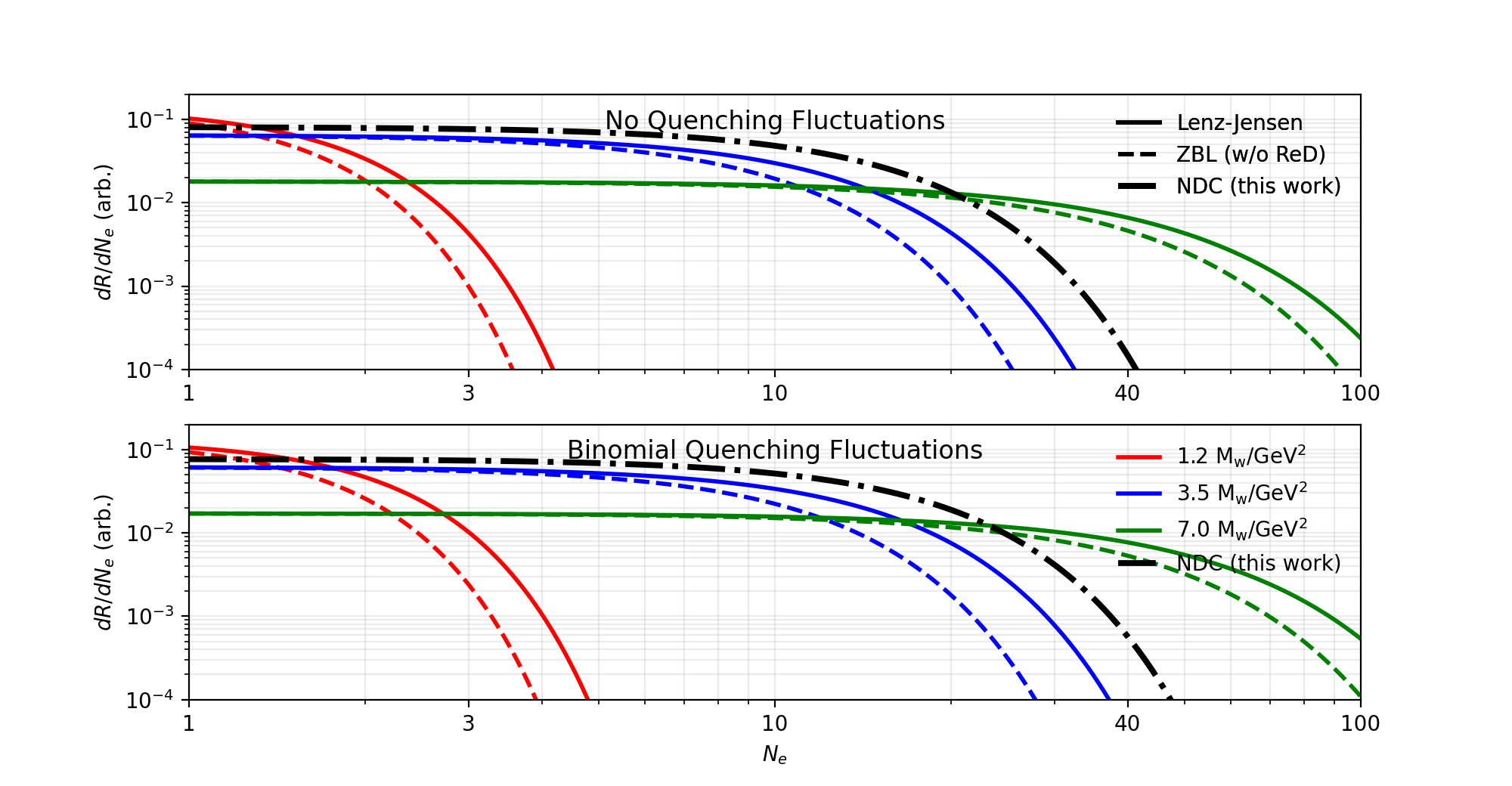}
	\caption{
		Probability density functions of the expected WIMP-induced ionization spectra in
		DarkSide-50 for WIMP masses of 1.2, 3.5, and 7.0~GeV/$c^2$, shown for three different
		screening-function models. The $f_q(E_{nr})$ response derived from the ZBL screening
		function corresponds to the previous fit excluding ReD data~\cite{DarkSide:2021bnz},
		while the $f_q(E_{nr})$ curve based on the Moli\`ere screening function is obtained
		in this work. In the upper panel, fluctuations in the nuclear recoil quenching
		process (NQ) are neglected, whereas in the lower panel, quenching fluctuations are
		described using a binomial distribution (QF).
	}
	\label{fig:wimps}
\end{figure}

The effect of the Lenz--Jensen screening-function-based $f_q(E_{nr})$, relative to the
previous ZBL-based model, on the predicted WIMP spectra is shown in
Fig.~\ref{fig:wimps} for several WIMP masses, assuming the DarkSide-50 detector
energy resolution~\cite{DarkSide-50:2022qzh}. Owing to the lack of a fully stochastic
description of the energy quenching mechanism for nuclear recoils, two cases are
considered: one in which quenching fluctuations are ignored (NQ), and another in which
binomial fluctuations between observable and unobservable quanta are included (QF).

The Lenz--Jensen screening function yields a larger ionization response at low recoil
energies compared to the ZBL screening function, thereby enhancing the likelihood that
events exceed the analysis thresholds. This leads to an improved projected sensitivity,
particularly for WIMPs with masses of order $\mathcal{O}(1~\mathrm{GeV}/c^2)$, which
generate nuclear recoils near the DarkSide experimental thresholds. We subsequently
recompute both the observed DarkSide-50 and the expected DarkSide-20k exclusion limits,
replacing the ZBL screening function with the Lenz--Jensen one, motivated by its superior
agreement with the complete calibration dataset. 
as it produces WIMP spectra that are nearly indistinguishable from those obtained using
the Moli\`ere model.

The 90\% confidence-level exclusion limits are obtained using a binned
profile-likelihood analysis, employing the same methodology and inputs—including
statistical and systematic uncertainties, background descriptions, and detector
resolutions—as in previous DarkSide-50~\cite{DarkSide-50:2022qzh} and
DarkSide-20k~\cite{DarkSide-20k:2024yfq} limit determinations. For DarkSide-50, the analysis
threshold is fixed at 4~$e^-$, with a dataset corresponding to an effective exposure of
approximately 12~ton~day after all selection criteria. For DarkSide-20k, the simulated
dataset assumes 10 years of operation, corresponding to an effective exposure of
342~ton~year and a 2~$e^-$ analysis threshold. The DarkSide-20k background model incorporates
spurious electron events, which dominate the rate in the few-$e^-$ region. Although their
physical origin remains not fully understood, they are observed to correlate with impurity
levels in liquid argon. Additional details can be found in
Refs.~\cite{DarkSide-50:2025umf, DarkSide-50:2022qzh, DarkSide-20k:2024yfq}.

The nuclear dielectric constant $\epsilon_n(q)$ provides an effective parameterization of additional nuclear polarization and screening effects at low momentum transfer. In the context of liquid argon ionization yield models, we implement $\epsilon_n(q)$ as a multiplicative correction to the Lenz--Jensen screening response, such that $Q_y(E_R)\rightarrow Q_y(E_R)\times \epsilon_n(q)$. This modification preserves agreement with ARIS, SCENE, and DS-50 data above $\sim$10~keV while enhancing the ionization yield at sub-keV energies, relevant for low-mass dark matter searches and S2-only analyses.

\begin{figure}
	\includegraphics[width=0.99\columnwidth]{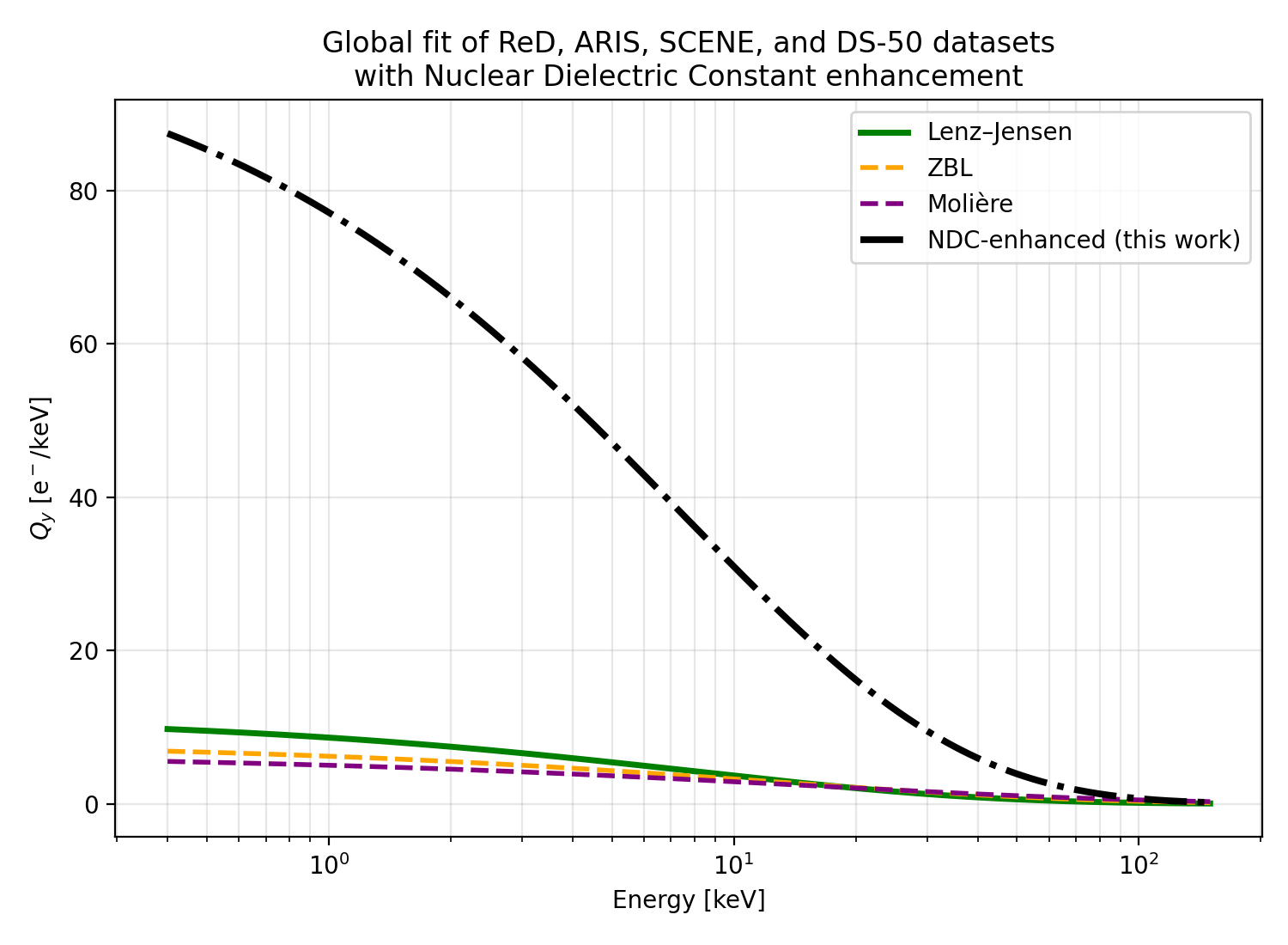}
	\caption{Simultaneous fit to the ReD, ARIS, SCENE, and DarkSide-50 as well the nuclear dielectric constant enhancement (black) datasets assuming the Lenz-Jensen screening function (green solid line). ReD data points are shown with both prior (gray) and posterior (red) uncertainties. The gray line and its corresponding uncertainty band represent the previous ionization model, based on the ZBL screening function and fitted without ReD data~\cite{DarkSide:2021bnz}. For comparison, the global fit including the ReD dataset was also performed using the screening functions of ZBL (orange dashed line) and Molière (purple dashed line). A curve labeled ``NDC-enhanced (this work)'' illustrates the effect of a nuclear dielectric constant $\epsilon_n(q)$ applied as a multiplicative correction to the Lenz--Jensen screening response at low momentum transfer, enhancing the predicted ionization yield below $\sim$3~keV without affecting the high-energy behavior constrained by DS-50 calibration data.
	} 
	\label{fig:qy}
\end{figure}

\begin{figure}
	\includegraphics[width=0.99\columnwidth]{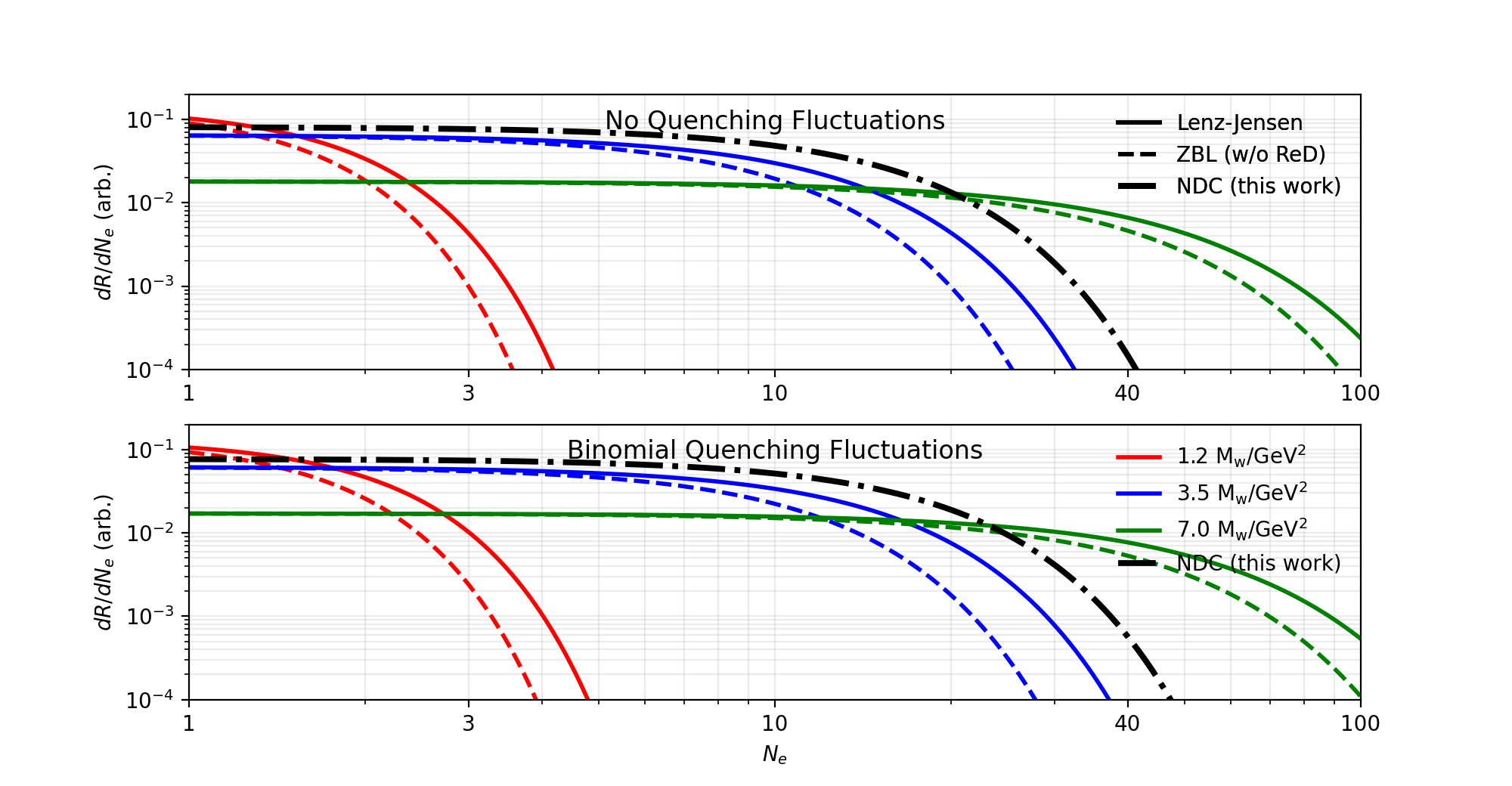}
	\caption{
		Probability density functions of the expected WIMP-induced ionization spectra in DarkSide-50 for WIMP masses of 1.2, 3.5, and 7.0~GeV/c$^2$, shown for three different SF models. The $f_q(E_{nr})$ response based on the ZBL SF corresponds to the previous fit without ReD data~\cite{DarkSide:2021bnz}, while the Molière SF-based $f_q(E_{nr})$ curve is from this work. In the top panel, no fluctuations are assumed in the nuclear recoil quenching process (NQ), whereas in the bottom panel, quenching fluctuations are modeled with a binomial distribution (QF).
	} 
	\label{fig:wimps}
\end{figure}


Liquid argon (LAr) time-projection chambers infer nuclear-recoil (NR) energies through the
ionization channel, commonly expressed as the ionization yield
\begin{equation}
	Q_y(E_R)\;\equiv\;\frac{N_e(E_R)}{E_R}\,,
	\label{eq:Qy_def}
\end{equation}
where $E_R$ is the NR energy and $N_e$ is the number of extracted ionization electrons
that generate the electroluminescence signal (S2). In global fits to
ReD/ARIS/SCENE/DarkSide-50 data, the low-energy behavior of $Q_y(E_R)$ is controlled by the
microscopic stopping and screening physics via a choice of screening function (e.g.\
Lenz--Jensen, ZBL, or Moli\`ere), together with a recombination prescription.

A convenient schematic factorization is
\begin{equation}
	N_e(E_R)\;=\;N_i(E_R)\,\bigl[1-r(E_R,E_d)\bigr]\,,
	\label{eq:Ne_recomb}
\end{equation}
where $E_d$ is the drift field, $r(E_R,E_d)$ is the recombination probability, and
$N_i(E_R)$ is the number of ion pairs initially created. The latter may be written as
\begin{equation}
	N_i(E_R)\;=\;\frac{E_R}{W}\,f_{\rm quanta}(E_R)\,,
	\label{eq:Ni_partition}
\end{equation}
with $W$ the effective work function in LAr and $f_{\rm quanta}(E_R)$ encoding the
fraction of NR energy that is converted into ionization/excitation quanta. In screening-based
global-fit models, the choice of screening function modifies $f_{\rm quanta}(E_R)$ at low
recoil energy, and therefore changes $Q_y(E_R)$.

\begin{figure}[t]
	\centering
	\includegraphics[width=\columnwidth]{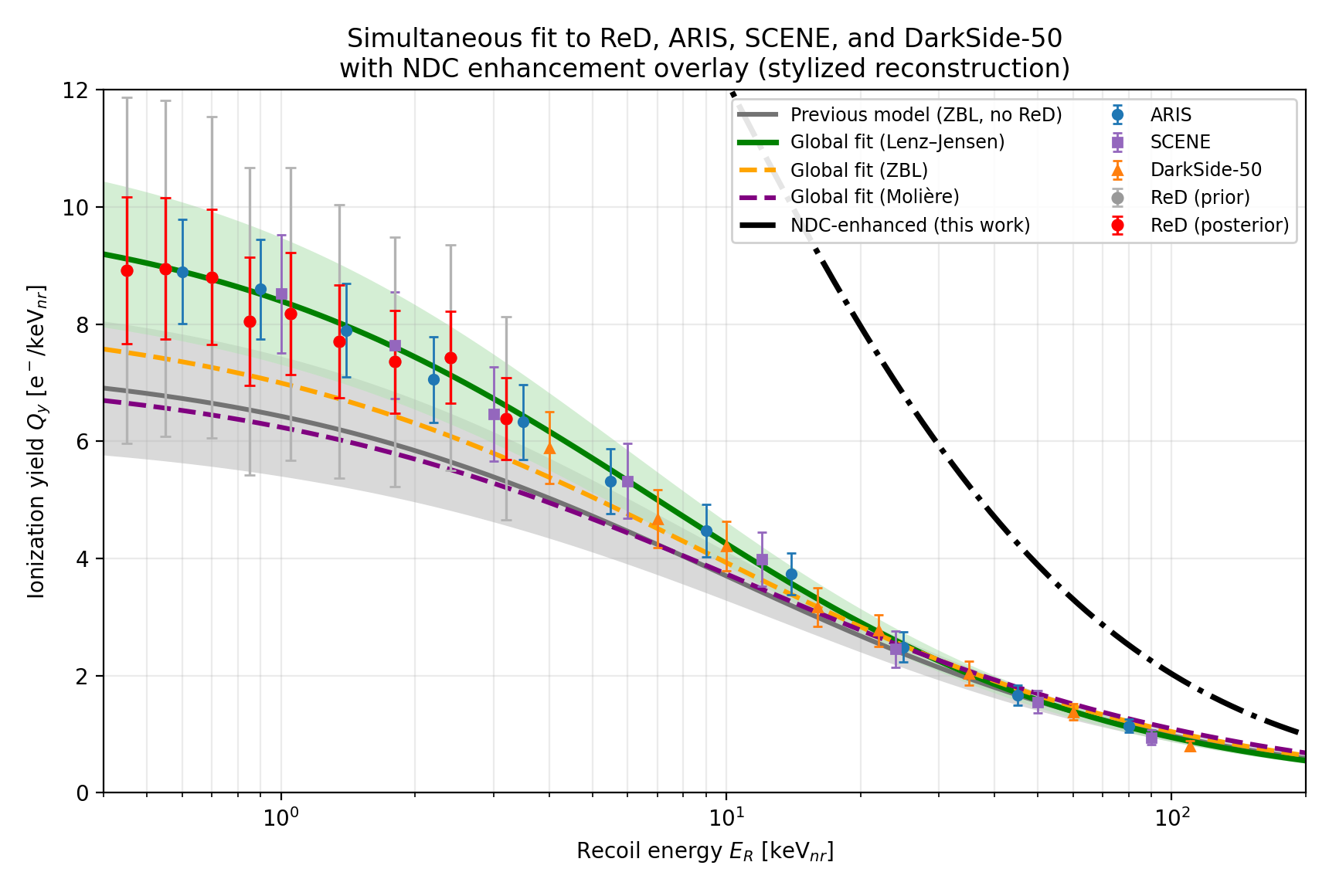}
	\caption{
		Stylized reconstruction of the simultaneous fit to ReD, ARIS, SCENE, and DarkSide-50
		ionization-yield datasets using different screening functions.
		The green solid curve represents the Lenz--Jensen (LJ) baseline fit; the orange and purple
		dashed curves show alternative screening functions (ZBL and Moli\`ere). The gray curve and band
		illustrate a previous ZBL-based model without ReD constraints. ReD points are shown with
		larger prior (gray) and reduced posterior (red) uncertainties.
		The black dot--dashed curve labeled ``NDC-enhanced (this work)'' overlays the baseline by
		applying the nuclear dielectric factor $\epsilon_n(q)$ through
		Eq.~\eqref{eq:Qy_NDC_insert}, enhancing $Q_y(E_R)$ at low recoil energies (small $q$) while
		leaving the high-energy behavior essentially unchanged.
	}
	\label{fig:Qy_fit_NDC}
\end{figure}

The nuclear dielectric constant (NDC) $\epsilon_n$ is introduced as a controlled,
low-momentum-transfer modification of the effective nuclear response. Defining the momentum
transfer
\begin{equation}
	q(E_R)\;=\;\sqrt{2m_N E_R}\,,
	\label{eq:q_of_ER}
\end{equation}
we adopt a phenomenological NDC factor
\begin{equation}
	\epsilon_n(q)\;=\;1+\frac{\alpha}{q^2+\beta}\,,
	\label{eq:epsn_model}
\end{equation}
with free parameters $(\alpha,\beta)$. Conceptually, $\epsilon_n(q)$ represents additional
screening/polarization/deformation effects that become increasingly relevant at small $q$.
To avoid double counting with the baseline screening description, we apply $\epsilon_n$ as
a multiplicative deformation of the baseline $Q_y(E_R)$ prediction:
\begin{equation}
	\boxed{
		Q_y^{\rm NDC}(E_R)\;=\;Q_y^{\rm base}(E_R)\,\epsilon_n\!\bigl(q(E_R)\bigr)
	}
	\label{eq:Qy_NDC_insert}
\end{equation}
(equivalently $N_e^{\rm NDC}(E_R)=N_e(E_R)\,\epsilon_n(q)$). In the context of the global fit
shown below, the baseline may be chosen as the Lenz--Jensen best-fit curve,
$Q_y^{\rm base}\equiv Q_y^{\rm LJ}$, and the NDC correction then enhances the ionization yield
primarily at sub-keV to few-keV recoil energies, while preserving the high-energy behavior
constrained by calibration data.

for implication for low-threshold (S2-only) analyses, since S2-only low-mass dark matter searches are effectively limited by a threshold on the
number of extracted electrons, an enhancement of $Q_y(E_R)$ at small $E_R$ increases the
probability that low-energy recoils exceed the few-electron threshold. In this sense, the
NDC acts as a physically motivated lever that can improve sensitivity to low-mass WIMPs by
re-weighting the low-$q$ part of the recoil response. See FIG \ref{fig:Qy_fit_NDC}

 \section{Qubit Detection: Detector design}
While standard transmon qubits (which operate at microwave frequencies, $\sim 5\text{ GHz}$) cannot "see" a 128 nm UV photon directly, we are using Superconducting Nanowire Single-Photon Detectors (SNSPDs) integrated onto the same chip as the qubit. These nanowires can detect a single UV photon and then send a signal to the qubit to be processed. 
To achieve a "zero-background" discovery of sub-GeV Dark Matter, the most advanced experimental architecture is the Monolithic Quantum Hybrid Calorimeter. This design integrates Superconducting Nanowire Single-Photon Detectors (SNSPDs) for light detection and a Qubit-based "Two-Chip" Phonon Sensor into a single cryogenic environment. This combination allows us to measure both the Scintillation (Light) and Recoil (Heat) of a single Dark Matter event with nearly 100\% efficiency. The device is constructed in a "flip-chip" geometry to isolate the quantum sensors from the massive Argon target while maintaining thermal contact. Chip A (The Target/Absorber): A high-purity crystalline substrate (like Sapphire) in direct contact with the Liquid Argon. The surface of this chip is patterned with an array of SNSPDs. Chip B (The Sensor): A superconducting chip containing the Transmon Qubit Arrays. The two chips are bonded using Indium bumps. These bumps act as "tunnels" for athermal phonons to travel from the target to the qubit. When a Dark Matter particle passes through the "empty space" of the Argon atom and hits a nucleus, it triggers a two-step quantum cascade: Step 1: The Optical Trigger (SNSPD), As the Argon excimers de-excite, they emit 128 nm UV photons.The SNSPDs on Chip A are held just below their critical current ($I_c$).A single UV photon hits the nanowire, creating a local "hotspot" that destroys superconductivity, resulting a massive voltage spike provides a nanosecond-accurate Time-Zero ($t_0$) for the event. Step 2: The Phonon Parity Flip (Qubit)Simultaneously, the nuclear recoil generates a burst of athermal phonons. The phonons propagate through the Argon/Sapphire interface, across the Indium bumps, and into the Qubit chip. For pair breaking, Phonons hit the Aluminum traces of the qubit, breaking Cooper pairs into Quasiparticles. So, for tunneling, These quasiparticles tunnel across the Josephson Junction of the Transmon, getting parity flip as Every tunneling event changes the charge parity of the qubit ($|e\rangle \leftrightarrow |o\rangle$). This results in that The qubit's resonant frequency shifts by a discrete amount ($\Delta \omega_q$), which we measure via a microwave readout. A valid Dark Matter candidate must satisfy the "Conservation of Parity" in the detector:

 \begin{equation}
N_{events} = \int (S_{SNSPD} \cap S_{Qubit}) dt
\end{equation}

Where the SNSPD signal $S$ must arrive within the Phonon Ballistic Time ($\approx 1\text{--}10\text{ }\mu\text{s}$) of the Qubit parity flip.

Also the detector adopts a flip-chip configuration composed of an upper {\it qubit chip} and a lower {\it carrier chip}, as illustrated in Figs.~\ref{fig:QPD_figure}(a)--\ref{fig:QPD_figure}(c). The qubit chip, fabricated on a sapphire substrate, each formed by a superconducting Al (aluminum)/Al$_2$O$_3$/Al Josephson junction and large-area Ta (tantalum) films galvanically coupled to the Al layer and functioning as capacitor pads. Dark matter scatters Argon nucleus and send phonon within the substrate of the qubit chip, generating a phonon that propagates from the substrate into the Ta films, where Cooper pairs are broken and quasiparticles (QPs) are produced. These QPs subsequently accumulate in the lower-gap Al region of the junction. Tunneling of QPs modifies the charge-parity state of the qubit and produces measurable signals.

More specifically, the parity-dependent Hamiltonian of the qubit is
\begin{equation}
	H = 4E_C\left(\hat{n} - n_g + \frac{P-1}{4}\right)^2 - E_J \cos\hat{\phi} ,
\end{equation}

where $E_C$ and $E_J$ denote the charging and Josephson energies, $\hat{n}$ and $n_g$ represent the Cooper-pair number operator and offset charge, $\hat{\phi}$ is the superconducting phase difference across the junction, and $P=+1$ ($-1$) labels the even (odd) charge-parity state. The qubit therefore exhibits two lowest-energy states, shown in Fig.~\ref{fig:QPD_figure}(e). Quasiparticle tunneling induces a frequency shift $|f_{\mathrm{e}}-f_{\mathrm{o}}|$, which is exploited to map charge-parity states onto qubit eigenstates using a Ramsey-based protocol~\cite{Rist2013,Serniak2018hot}. The qubit array thus operates as a collection of QPDs, capable of resolving individual QP-tunneling events and sensitive to low-energy single phonons. Experimentally, this technique has already achieved time resolutions of several microseconds~\cite{Li:2024dpf,Rist2013}. In addition, fidelities of $99.9\%$ have been demonstrated for both single-qubit gates and readout~\cite{Wang:2024rjw,marxer2025above}. Accordingly, we assume a charge-parity detection fidelity of $\mathcal{F}=95\%$ in this work. Each qubit is also equipped with an auxiliary gate line to compensate offset-charge drift~\cite{Christensen2019Anomalous,wilen2021correlated}, enabling stable detector operation over periods of weeks to months.

Finally, the detector is packaged and operated at the base temperature of a dilution refrigerator near $10~\mathrm{mK}$ (Fig.~\ref{fig:QPD_figure}(d)), with the entire apparatus enclosed within gamma and neutron shielding at CJPL. The dominant low-energy background arises from tunneling induced by residual QPs. Near-gap infrared photons originating from higher-temperature refrigerator stages can break Cooper pairs and enhance the residual QP density $n_{\rm res}$, primarily via the coaxial cabling~\cite{Liu:2022aez}. To mitigate this effect, light filtering is applied to the wiring, and a compact sample enclosure coated with infrared-absorbing material is used to improve shielding against infrared and stray photons. This approach is expected to suppress the tunneling rate to approximately 0.1~Hz~\cite{Connolly:2023gww,pan2022engineering}, corresponding to a residual QP density of $n_{\rm res}\sim 10^{-4}~\mu{\rm m}^{-3}$. Conservatively, we adopt $n_{\rm res}=10^{-3}~\mu{\rm m}^{-3}$ as a benchmark value.

This "Two-Chip" hybrid model represents the ultimate limit of sensitivity. It uses the SNSPD to overcome the "timing noise" of superconducting circuits and the Qubit to overcome the "energy threshold" of silicon-based sensors.

As demonstrated below, the detector exhibits several distinctive advantages: (1) The suspended geometry forces DM-induced phonons to dissipate into the environment (carrier chip) exclusively through four In (indium) pillars located at the corners. Simulations indicate that more than 80\% of the phonon energy is absorbed by the qubits. (2) The Ta--Al structure concentrates QPs within the Al trap, significantly enhancing the QP tunneling rate. (3) The detector is sensitive to single-QP tunneling and low-energy individual phonons. The simulated detection efficiency approaches $100\%$ in the sub-eV regime, in contrast to the $\mathcal{O}({\rm eV})$ thresholds typical of TES- and skipper-CCD-based dark-matter experiments~\cite{SuperCDMS:2024yiv,SENSEI:2023zdf}. An energy resolution of approximately 50\% at 100~meV is also achieved, enabling rejection of higher-energy backgrounds. (4) The detector mass can be scaled with minimal loss of detection efficiency. We propose multiple geometries for the qubit chip, including a ``thin chip'' of 0.82~g ($21.2\times22.6\times0.43$~mm), typical of quantum-computing devices~\cite{Li:2024dpf}, and a ``thick chip'' of 38~g ($21.2\times22.6\times20$~mm). The final experimental phase targets kilogram-scale masses using arrays of thick chips.

\section{DM interaction rate}
Light dark matter $\chi$ can interact with electrons in the target material and produce a sub-eV phonon excitation, with a rate determined by the intrinsic response properties of the medium. The differential interaction rate $\mathrm{d}R/\mathrm{d}\omega$ is evaluated using the \texttt{PhonoDark}~\cite{Trickle:2020oki} framework. As required inputs, we compute the phonon dispersion relations, equilibrium ionic configurations, total potential energy, and effective ionic charges employing first-principles density functional theory calculations implemented with the \texttt{VASP} package~\cite{kresse1996efficient,kresse1996efficiency}.

The same detector configuration is also sensitive to dark matter absorption processes. The absorption rate of a dark photon within the sapphire target is proportional to the imaginary component of the inverse dielectric function~\cite{Knapen:2021bwg,Berlin:2023ppd,Hochberg:2016sqx,Griffin:2018bjn,Mitridate:2021ctr,Mitridate:2023izi}, which is obtained using the \texttt{darkELF} toolkit~\cite{Knapen:2021bwg}. The deposited energy is approximately monochromatic and manifests as a phonon with energy equal to the dark-photon mass. Axions may be detected in an analogous manner through the application of an external magnetic field.

For both dark matter scattering and absorption channels, the predicted number of signal events in a detector with mass $M$ operated over an exposure time $\mathcal{T}$ is given by
\begin{equation}
	N_{\rm sig}=\int \mathrm{d}\omega\, M\mathcal{T}\,\frac{\mathrm{d}R}{\mathrm{d}\omega}
	\int \mathrm{d}\omega_{\rm rec}\,\varepsilon(\omega)\,f_{\rm r}(\omega_{\rm rec},\omega)\,,
	\label{eq:Nsig}
\end{equation}
where $\omega_{\rm rec}$ denotes the reconstructed phonon energy, while $\varepsilon$ and $f_{\rm r}$ represent the detector efficiency and energy-resolution response functions.

\section{Phonon transport and quasiparticle tunneling rate}
 We perform simulations of the propagation and complete phonon dynamics generated by dark matter interactions using the \texttt{G4CMP} framework~\cite{Kelsey:2023eax}, based on the optimized geometry of the qubit-based detector. Phonons are injected uniformly throughout the substrate with energy $\omega$, and their initial polarization states are randomly drawn according to the partition fractions among the allowed polarization modes. During propagation, phonons scatter from lattice imperfections such as isotopic disorder, leading to changes in both propagation direction and polarization at a rate proportional to $B\nu^4$, while longitudinal modes may undergo anharmonic decay with a rate $A\nu^5$. When phonons reach the sapphire--Ta interface, they transmit into the Ta absorber with a probability of 94\%, as determined using the Acoustic Mismatch Model~\cite{Swartz1989rmp}, and once absorbed they are unlikely to escape back into the substrate~\cite{Kelsey:2023eax}. Phonons that are not absorbed are reflected under the assumption of specular reflection, consistent with previous investigations~\cite{Martinez:2018ezx}. Additional phonon losses may occur through excitation of surface modes when interacting with non-ideal substrate boundaries. Following the CDMS treatment~\cite{CDMSsimulation}, we assume a single-pass surface loss probability of 0.1\%.
 
 Figure~\ref{fig:QPD_figure}(f) displays the fraction of phonon energy absorbed by the qubits, rather than the In pillars, as a function of the $x$--$y$ position of the dark matter interaction. For the multi-qubit configuration, the average absorbed fraction exceeds 80\% over most of the substrate area, in contrast to values of $\lesssim 10\%$ reported for comparable quantum devices~\cite{Martinez:2018ezx,Linehan:2025suv}. In regions close to the pillars, the absorption fraction decreases to below 30\%.
 
 Phonons entering the Ta absorber with energies $E_{\rm ph}>2\Delta_{\rm abs}$, corresponding to twice the superconducting gap of the absorber, are capable of breaking Cooper pairs through the Kaplan cascade mechanism~\cite{Kaplan1976prb}, producing quasiparticles with energies approaching $\Delta_{\rm abs}$. The mean fraction of phonon energy converted into quasiparticles is described by $\eta_{\rm pb}$, which accounts for losses to sub-gap phonon production. We adopt $\eta_{\rm pb}=0.6$, consistent with values reported in the literature~\cite{guruswamy2014quasiparticle,kozorezov2000quasiparticle}.
 
 Because the superconducting gap of the Al trap $\Delta_{\rm trap}$ is significantly smaller than that of the Ta absorber, quasiparticles subsequently diffuse from the absorber into the trap. We define the trapping efficiency $\eta_{\rm trap}$ as the fraction of quasiparticles generated in the absorber that successfully enter and remain confined within the trap. By solving the quasiparticle diffusion equation between the absorber and trap, we find $\eta_{\rm trap}\sim 0.7$ for our detector configuration. The resulting quasiparticle number density in the trap due to absorption of phonon energy $E_{\rm ph}$ is therefore given by $n_{\rm qp}=\eta_{\rm pb}\eta_{\rm trap}E_{\rm ph}/(V_{\rm trap}\Delta_{\rm trap})$, where $V_{\rm trap}$ denotes the trap volume.
 
 Following their creation, quasiparticles undergo exponential depletion via recombination on a characteristic timescale $\tau_{\rm qp}$ of order a few milliseconds~\cite{barends2009enhancement,wang2014measurement}. To remain conservative, we take $\tau_{\rm qp}=1$~ms in this analysis. The actual quasiparticle population may fluctuate due to correlations in the creation process, characterized by a Fano factor $F\sim0.2$~\cite{verhoeve2002superconducting}, leading to a normal distribution with variance $\sigma=\sqrt{F n_{\rm qp} V_{\rm trap}}$~\cite{fano1947ionization}. The quasiparticle tunneling rate is calculated as~\cite{palmer2007steady,shaw2008kinetics,shaw2009quantum,lutchyn2007kinetics}
 \begin{equation}
 	\Gamma_{\rm tun}(t)\simeq \frac{16E_J k_B T}{h\mathcal{N}_{\rm qp}\Delta_{\rm trap}}\,n_{\rm qp}e^{-t/\tau_{\rm qp}}\,,
 \end{equation}
 in the low-temperature regime $T\ll\Delta_{\rm trap}$, where $\mathcal{N}_{\rm qp}$ denotes the number of available quasiparticle states in the trap. The background tunneling rate arising from a nonzero residual quasiparticle density $n_{\rm res}$ is obtained by an analogous expression, replacing $n_{\rm qp}e^{-t/\tau_{\rm qp}}$ with $n_{\rm res}$.

\section{Detector resolution and efficiency}
The total number of tunnel events at a qubit during a single measurement consists of the sum of phonon-driven tunnelings $N_{s}$ and residual quasiparticle (QP)-induced tunnelings $N_{0}$. These contributions are sampled from a Poisson distribution while simultaneously accounting for the detector readout fidelity.

\begin{figure}[!htb]
	\centering
	\includegraphics[width=\columnwidth]{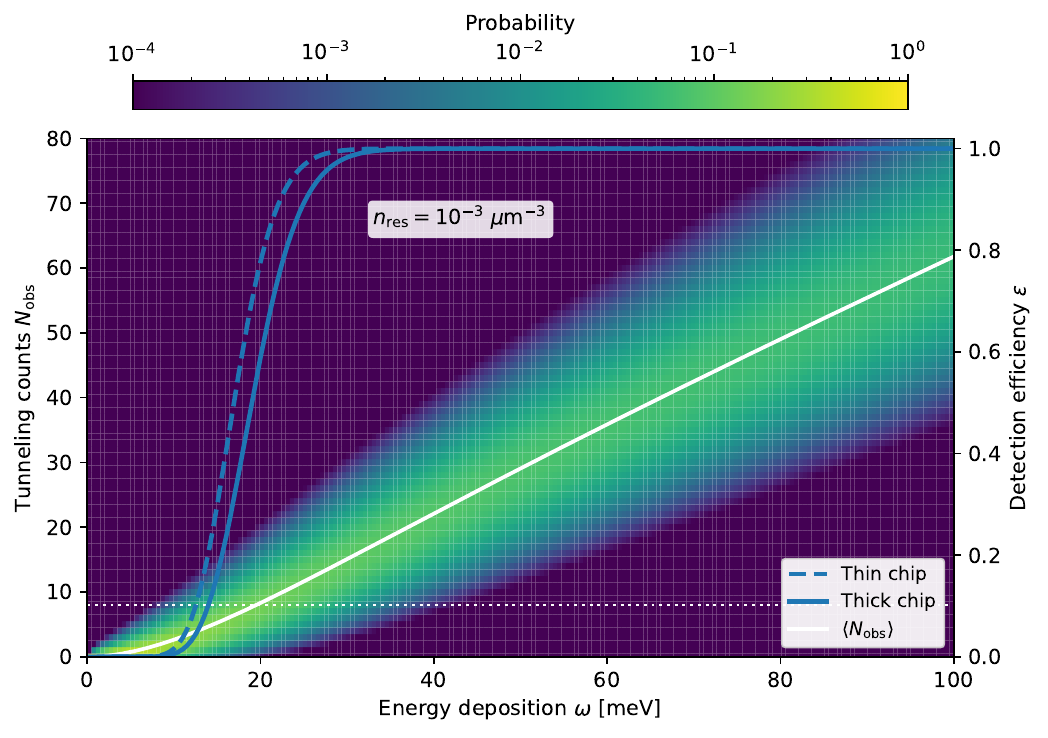}
	\includegraphics[width=\columnwidth]{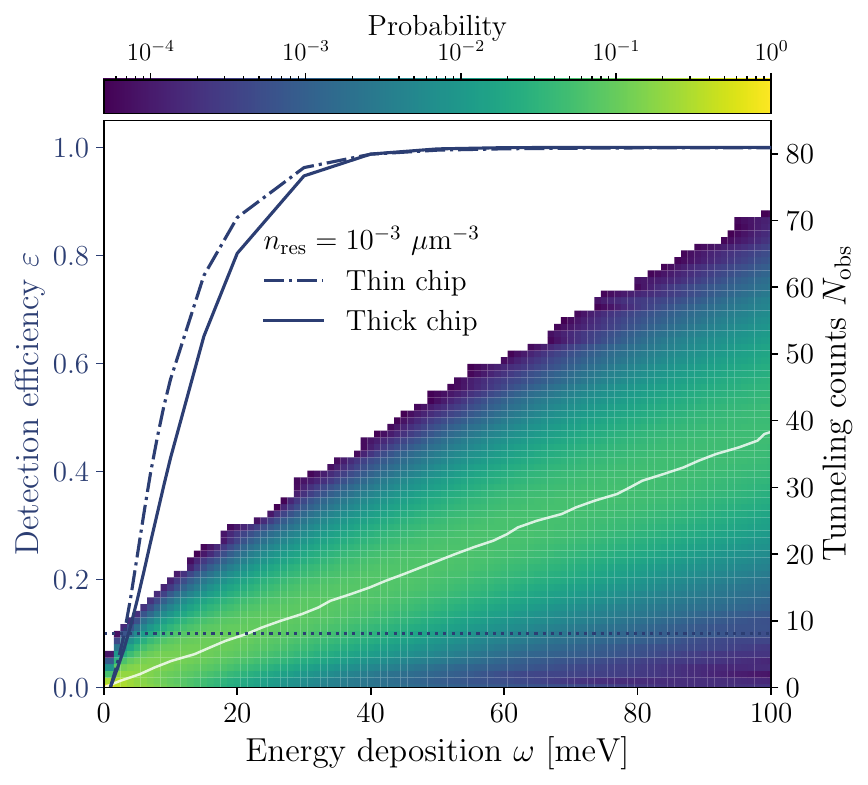}
	\caption{The blue curves indicate the detection efficiency as a function of the deposited energy from DM interactions for a thin-chip (dash-dotted) and a thick-chip (solid) detector configuration. The dotted line denotes the 10\% efficiency level. The color scale represents the probability distribution of tunneling counts across all qubits within a single measurement time window, $N_{\rm obs}$, for a thin chip as a function of deposited energy, while the white curve shows the reconstructed energy corresponding to the maximum-likelihood estimate as a function of $N_{\rm obs}$.}
	\label{fig:efficiency}
\end{figure}

The phonon energy $\omega$ generated by a DM interaction is inferred from the observed total number of tunnelings $N_{\rm obs}$ across all qubits within a 1.5~ms measurement window.
The reconstructed energy is determined using a maximum-likelihood approach, $\omega_{\rm rec} = {\max}_{\omega}\,\mathcal{L}(N_{\rm obs}|\omega)$, where the likelihood function $\mathcal{L}$ is derived from detector simulations, as illustrated in Fig.~\ref{fig:efficiency}. The energy resolution, defined as the ratio between the full width at half maximum (FWHM) and $\omega_{\rm rec}$, improves with increasing phonon energy and reaches approximately 50\% at $\omega=100$~meV. The resolution function $f_r$ is modeled as a Gaussian distribution with $\sigma=\textrm{FWHM}/2.355$. As a result, high-energy background events outside the signal window can be efficiently rejected.

We next determine the signal detection efficiency in the presence of background tunnelings arising from residual QPs. A signal is identified by requiring that the total number of tunnelings observed for $N_{\rm qb}=96$ qubits in a given measurement $j$ deviates from the expected background level $N_b=\mathcal{F}N_{0}N_{\rm qb}$ (background-only hypothesis) by at least $3\sigma$ confidence level, assuming a $\chi^2$ distribution. This corresponds to the condition $\Sigma_{N_{\rm obs}=N_j}^{\infty} \mathrm{Poisson}(N_{\rm obs},\mu=N_b)<0.0027$. The detection efficiency is then evaluated as the fraction of simulated events satisfying this criterion.

\begin{figure}[!htb]
	\centering
	\includegraphics[width=\columnwidth]{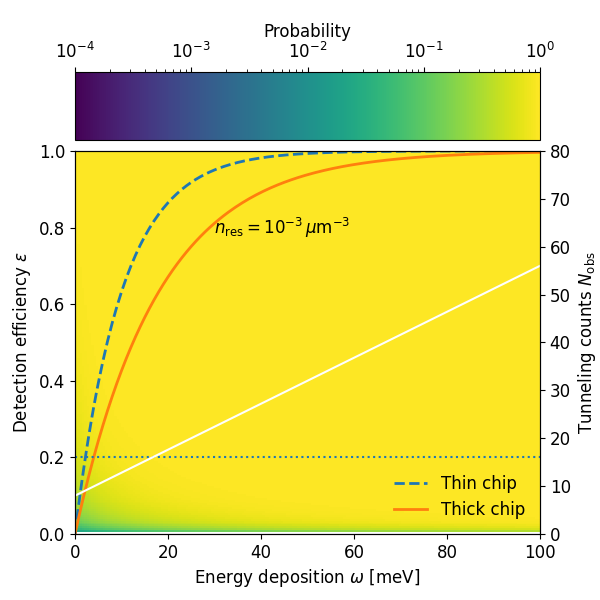}
	\caption{The the detection efficiency as a function of the deposited energy from DM interactions for a thin-chip (dash-dotted) and a thick-chip (solid) detector configuration with SNSPD based on simulations}
	\label{fig:efficiencySNSPD}
\end{figure}

The resulting detection efficiency $\varepsilon$ is shown in Fig.~\ref{fig:efficiency}, where it rapidly approaches unity as the phonon energy increases. A more than 10\% efficiency is achieved at $\omega\sim3$~meV for both thin- and thick-chip geometries, representing several orders of magnitude improvement over the thresholds of conventional direct-detection experiments. It is a good enhancement based on Ref.~\cite{li2025novellightdarkmatter}. The thick-chip configuration exhibits only a modest reduction in efficiency relative to the thin chip. Although phonons propagate over longer distances in the thicker substrate and are more likely to down-convert into lower-energy modes below the pair-breaking threshold, phonons at the few-meV scale possess mean free paths exceeding the dimensions of the thick chip, thereby suppressing efficient decay. But it is improved with SNSPD as shown in Fig.~\ref{fig:efficiencySNSPD}

The Hybrid SiPM-SNSPD-Qubit Detector provides a significant advancement over standalone qubit sensors by integrating multi-channel coincidence to achieve a "background-free" search environment across a broad energy spectrum. 
\begin{figure}[!htb]
	\centering
	\includegraphics[width=\columnwidth]{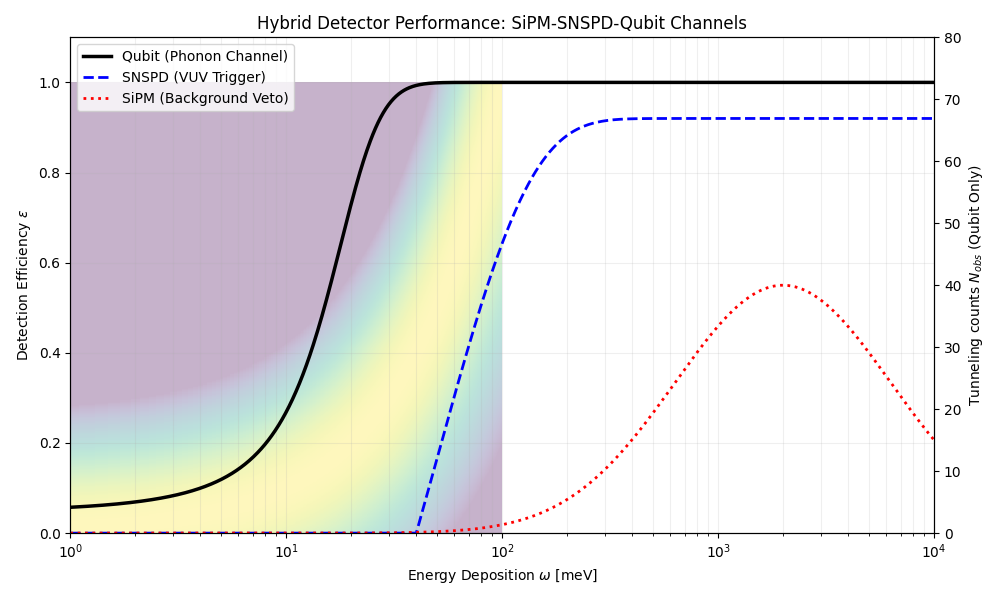}
	\caption{the detection efficiency as a function of the deposited energy from DM interactions for a Hybrid detector configuration.}
	\label{fig:probability2}
\end{figure}
While the standalone qubit array, as shown in the primary performance plot, is highly efficient at resolving sub-meV athermal phonons with a 10\% efficiency threshold at approximately 3 meV, it is inherently limited by a narrow dynamic range and vulnerability to residual quasiparticle noise ($n_{\text{res}} = 10^{-3} \mu\text{m}^{-3}$). In contrast, the hybrid system utilizes a Superconducting Nanowire Single-Photon Detector (SNSPD) as a high-precision VUV trigger to establish an event $t_0$, effectively filtering out random parity flips caused by environmental noise. Furthermore, the integration of a Silicon Photomultiplier (SiPM) provides a critical high-energy veto channel that identifies and rejects cosmic muons and radioactive backgrounds (such as $^{39}\text{Ar}$) that would otherwise saturate the sensitive qubit sensors. By extending the detection range from the meV phonon regime to the eV photon regime (as illustrated by the divergent efficiency curves in the hybrid performance plot \ref{fig:probability2}), the system ensures that every interaction from $1 \text{ meV}$ to $10,000 \text{ meV}$ is characterized with high fidelity, allowing for the exploration of light dark matter candidates well below the classical ionization floor.

\begin{figure*}[!htb]
	\centering
	\includegraphics[width=\columnwidth]{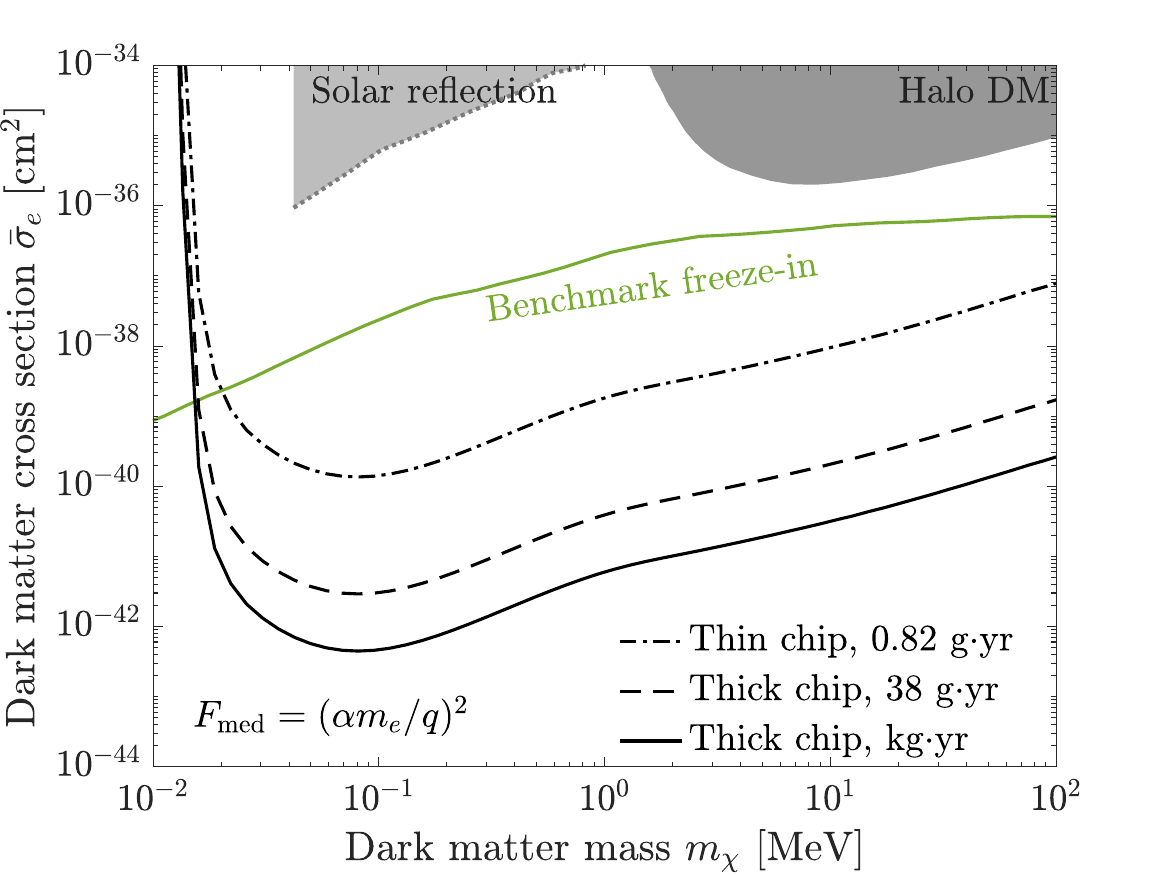}
	\includegraphics[width=\columnwidth]{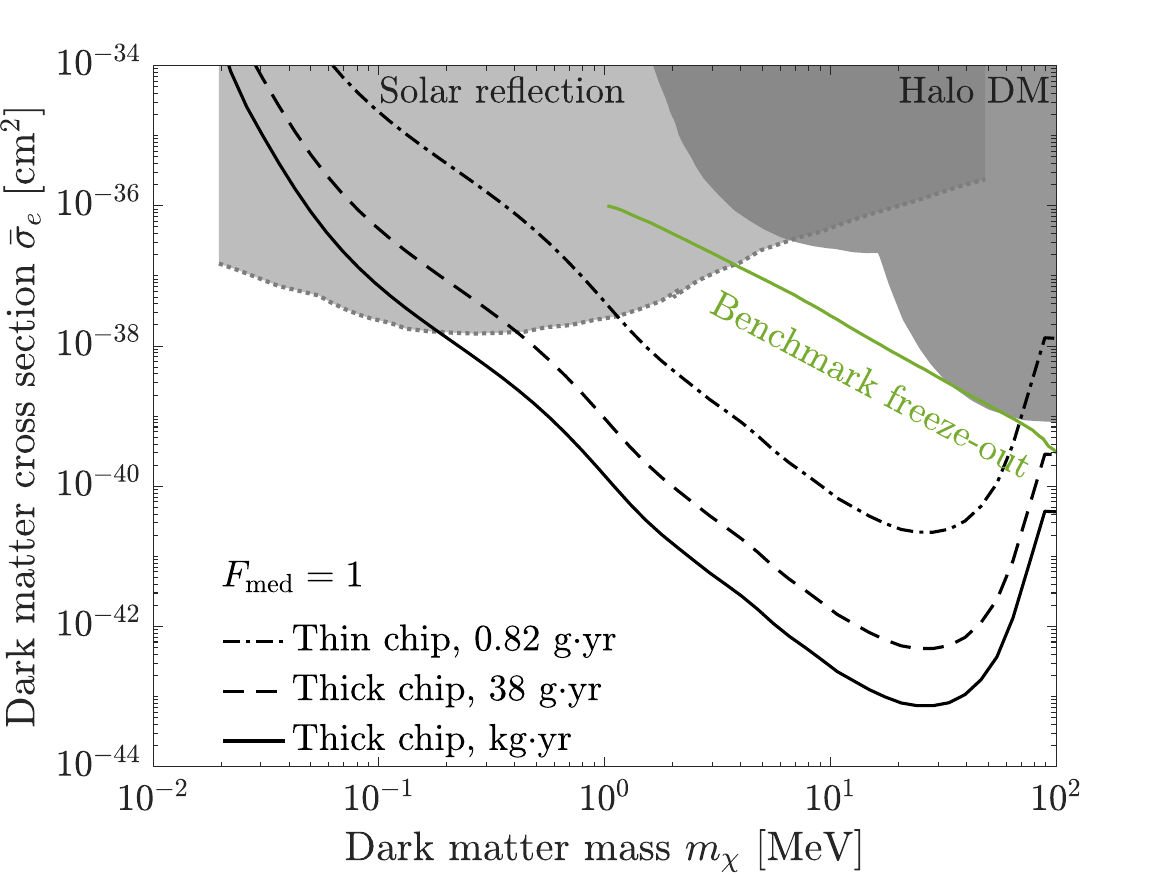}
	\includegraphics[width=\columnwidth]{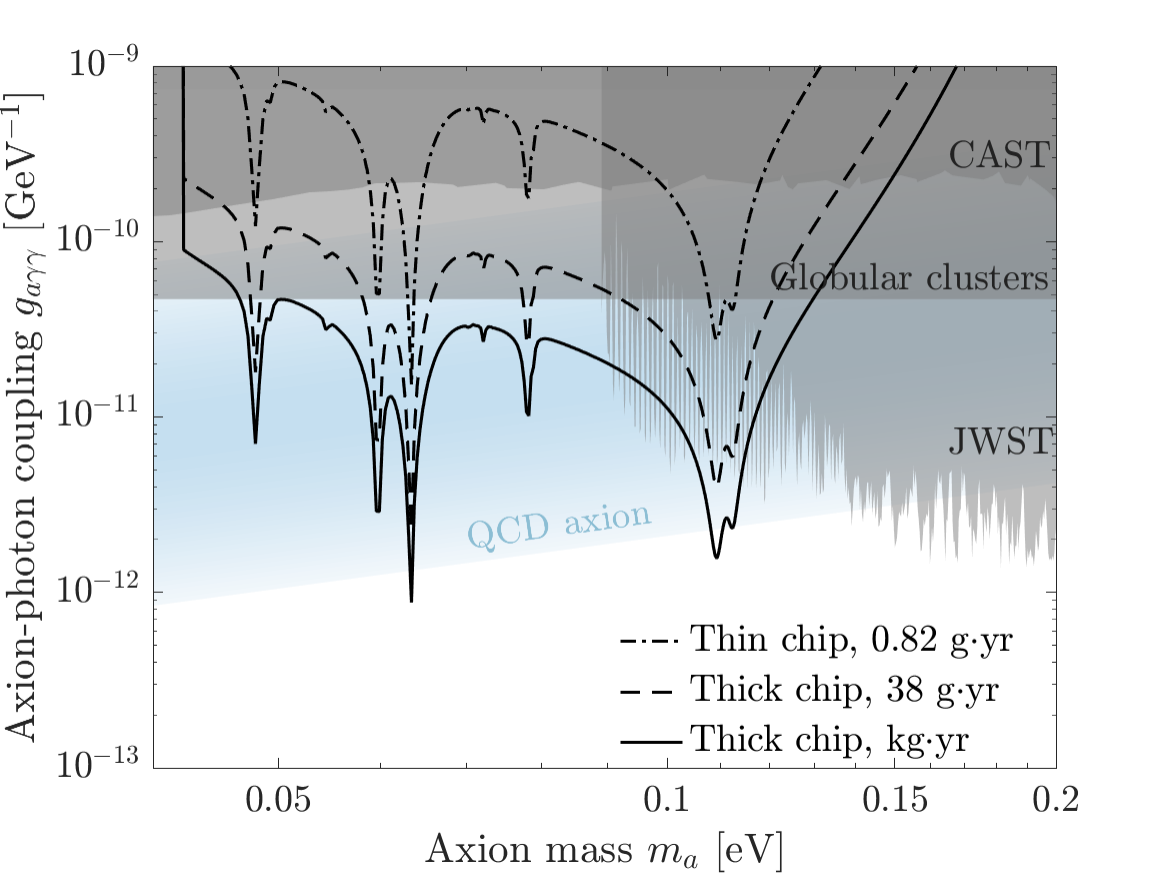}
	\includegraphics[width=\columnwidth]{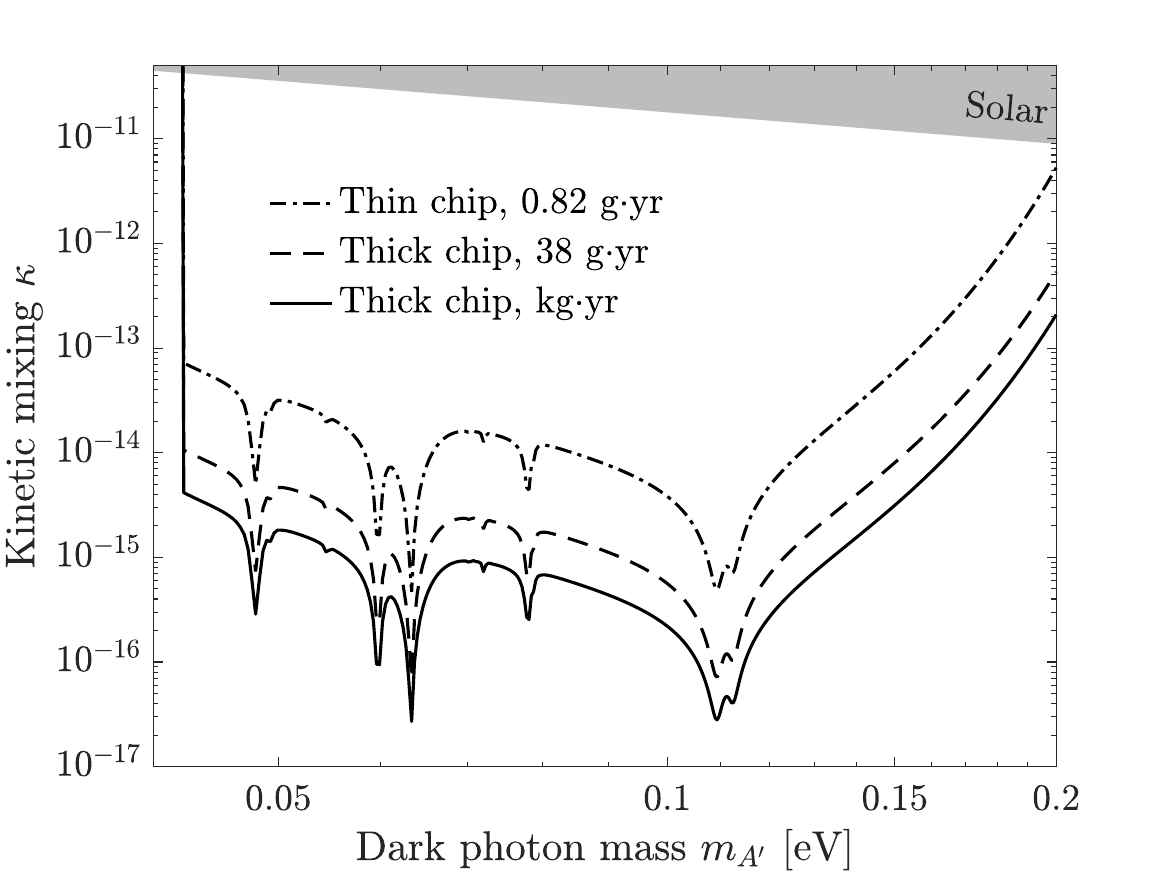}
	\caption{\textit{Upper:} Projected 90\% C.L. reach for the DM--electron scattering cross section. The black curves
		represent various detector exposures achievable with the setup developed in this study.
		Current bounds from direct detection experiments~\cite{SENSEI:2023zdf,DAMIC-M:2023gxo,
			DarkSide:2022knj,XENON:2019gfn,EDELWEISS:2020fxc,PandaX-II:2021nsg,SuperCDMS:2024yiv},
		together with constraints from DM acceleration and solar-electron reflection~\cite{An:2021qdl,
			Emken:2024nox}, are displayed for comparison. The green curves denote the cross sections
		required to reproduce the observed DM relic abundance under representative freeze-in
		mechanisms~\cite{Essig:2011nj,Chu:2011be,Essig:2015cda,Dvorkin:2019zdi} for light mediators
		(left) and freeze-out scenarios~\cite{Essig:2015cda,Boehm:2003hm,Lin:2011gj,Hochberg:2014dra,
			DAgnolo:2019zkf,Kahn:2021ttr} for heavy mediators (right). \textit{Lower:} Projected 90\%
		C.L. reach for DM absorption. For axions, existing constraints on the axion--photon coupling
		$g_{a\gamma\gamma}$ from CAST~\cite{CAST:2017uph,CAST:2024eil}, globular clusters~\cite{Ayala:2014pea,
			Dolan:2022kul}, and JWST~\cite{Pinetti:2025owq} are included (left). The blue region marks the
		parameter space of the QCD axion~\cite{Peccei:1977ur,Peccei:1977hh,Wilczek:1977pj,Weinberg:1975ui,
			AxionLimits}. For dark photons, restrictions on the kinetic mixing parameter $\kappa$ from solar
		cooling~\cite{Vinyoles:2015aba,An:2013yfc,Li:2023vpv} are displayed (right)
		 [based on \cite{li2025novellightdarkmatter} for constrains].}
	\label{fig:Al2O3sensitivity}
\end{figure*}

\section{Sensitivity and background}
We obtain the projected constraints on dark matter parameters by imposing that the expected number of signal events from DM interactions, evaluated using Eq.~\eqref{eq:Nsig}, does not exceed the anticipated background at the 90\% confidence level. Our analysis is restricted to the single-phonon regime, corresponding to recoil energies $\omega_{\rm rec}\leq 100$~meV, while contributions from multi-phonon excitations and ionization channels are deferred to future studies.

We consider two categories of background sources---tunneling processes induced by residual quasiparticles and soft scattering of energetic photons, where the former is incorporated into the overall detection efficiency. Contributions from cosmic rays and radioactive decays typically deposit energies well above the keV scale and lie far outside the signal region; these backgrounds are largely mitigated by the overburden and passive shielding at CJPL. In contrast, high-energy photons generated within the refrigerator can undergo soft scattering and produce energy depositions below 100~meV, rendering them indistinguishable from dark matter interactions on an event-by-event basis. We follow the background modeling presented in~\cite{Berghaus:2021wrp}, adopting the reference photon spectrum measured by the EDELWEISS Collaboration~\cite{EDELWEISS:2018tde}. This corresponds to employing shielding composed of 20~cm of lead and 50~cm of polyethylene to suppress gamma rays and neutrons in our experimental configuration. The resulting background rate is estimated to be 5.27~kg$^{-1}\cdot$yr$^{-1}$, which we take as the benchmark value throughout this work.

Furthermore, mechanical stress arising from differential thermal contraction of detector components during cooldown can generate phonon bursts in the qubit-based detector~\cite{anthony2024stress}. Such stress-related backgrounds decay exponentially with a characteristic time scale of approximately 6--10 days~\cite{anthony2024stress,Yelton:2025wsy} and can be effectively removed by delaying data acquisition sufficiently after cooldown. Consequently, these contributions are neglected in our sensitivity estimates.

We emphasize that the background level in the sub-100~meV energy window is currently poorly constrained experimentally and could deviate significantly from the assumed benchmark. The projected sensitivity therefore scales directly with the true background level realized in the experiment.

The projected reach for the DM--electron scattering cross section is presented in the upper panels of Fig.~\ref{fig:Al2O3sensitivity}. For both light and heavy mediators, the anticipated limits from a single thin-chip detector with an exposure of 0.82~g$\cdot$yr exceed existing bounds by several orders of magnitude, including constraints from direct detection of halo dark matter~\cite{SENSEI:2023zdf,DAMIC-M:2023gxo,DarkSide:2022knj,XENON:2019gfn} as well as limits derived from solar-reflected dark matter~\cite{An:2021qdl,Emken:2024nox}. This sensitivity enables comprehensive tests of benchmark freeze-in~\cite{Essig:2011nj,Chu:2011be,Essig:2015cda,Dvorkin:2019zdi} and freeze-out~\cite{Essig:2015cda,Boehm:2003hm,Lin:2011gj,Hochberg:2014dra,DAgnolo:2019zkf,Kahn:2021ttr} production mechanisms.

We additionally present the projected sensitivity to dark-photon kinetic mixing, which improves upon current bounds from solar cooling~\cite{Vinyoles:2015aba,An:2013yfc,Li:2023vpv} by approximately 4--5 orders of magnitude using a single thin qubit chip. An analogous detection strategy may be applied to axion dark matter in the presence of a strong external magnetic field $B_0$, the experimental viability of which has been demonstrated recently in~\cite{Gunzler:2025spt}. With one year of operation, a single thin chip can already exclude unexplored regions of parameter space for $B_0=10$~T, while exposures at the kg$\cdot$yr scale can strengthen existing constraints from globular clusters by up to roughly two orders of magnitude.

\section{Conclusions and discussions}
We demonstrate the performance of a newly developed cryogenic detector based on a qubit array. While our analysis concentrates on sensitivity to dark-matter–electron scattering and dark-matter absorption, comparable sensitivity and mass reach can also be obtained for dark matter interacting with nucleons~\cite{Trickle:2019nya}.

In this work, we reconstruct the deposited phonon energy using only the total number of detected tunneling events. In a multi-qubit configuration, the spatial arrangement of tunneling qubits together with timing correlations may further enhance energy and position reconstruction, as well as signal–background discrimination—since phonon-induced tunneling events decay rapidly in time, whereas residual tunneling processes remain approximately time-independent. In addition, lowering $\Delta_{\rm trap}$ could further increase detection efficiency, while modifying the substrate material would alter the detector response to different dark-matter interaction models~\cite{Griffin:2019mvc,Campbell-Deem:2022fqm}. A comprehensive investigation of these potential improvements is left for future studies.

We consider two primary background sources—tunneling events induced by residual quasiparticles and soft scattering of high-energy photons—with the former incorporated into the detection efficiency. Cosmic rays and radioactive decays generally produce energy depositions well above the keV scale and outside the signal region, and are largely suppressed by the surrounding rock and shielding at CJPL. However, energetic photons generated within the refrigerator can scatter at low energies and deposit less than 100~meV, making them indistinguishable from dark-matter interactions on an event-by-event basis. We follow the background estimation of Ref.~\cite{Berghaus:2021wrp}, adopting the benchmark photon spectrum measured by the EDELWEISS Collaboration~\cite{EDELWEISS:2018tde}. This corresponds to implementing 20~cm of lead and 50~cm of polyethylene shielding against gamma rays and neutrons in our setup. The resulting background rate is estimated to be 5.27~kg$^{-1}\!\cdot\!$yr$^{-1}$, which we take as a reference value throughout this analysis.

Additionally, stress released due to mismatched thermal contraction among detector materials during cooldown may generate phonon bursts in the qubit detector~\cite{anthony2024stress}. Such stress-related backgrounds decay exponentially with a characteristic timescale of approximately 6–10 days~\cite{anthony2024stress,Yelton:2025wsy} and can be mitigated by allowing sufficient waiting time prior to data acquisition. Consequently, we neglect these effects in our sensitivity calculations.

We emphasize that the background level below 100~meV has not yet been precisely measured and could deviate significantly from current expectations. The experimental sensitivity is therefore expected to scale proportionally with the true background rate.

The projected sensitivity to the dark-matter–electron scattering cross section is shown in the upper panels of Fig.~\ref{fig:Al2O3sensitivity}. For both light and heavy mediators, the anticipated limits from a single thin-chip detector with an exposure of 0.82~g$\!\cdot\!$yr exceed existing constraints by several orders of magnitude, including those from direct detection of halo dark matter~\cite{SENSEI:2023zdf,DAMIC-M:2023gxo,DarkSide:2022knj,XENON:2019gfn} and from dark matter reflected by the Sun~\cite{An:2021qdl,Emken:2024nox}. This sensitivity enables a detailed exploration of benchmark freeze-in~\cite{Essig:2011nj,Chu:2011be,Essig:2015cda,Dvorkin:2019zdi} and freeze-out~\cite{Essig:2015cda,Boehm:2003hm,Lin:2011gj,Hochberg:2014dra,DAgnolo:2019zkf,Kahn:2021ttr} scenarios.

We also present the sensitivity to dark-photon kinetic mixing, which improves upon existing limits from solar cooling~\cite{Vinyoles:2015aba,An:2013yfc,Li:2023vpv} by approximately four to five orders of magnitude using a single thin qubit chip. An analogous strategy can be employed to search for axion dark matter in the presence of a strong magnetic field $B_0$, the feasibility of which has recently been demonstrated experimentally~\cite{Gunzler:2025spt}. With one year of operation, a single thin chip can already exclude new regions of parameter space for $B_0=10$~T, while a kg$\!\cdot\!$yr exposure could improve current globular-cluster constraints by up to two orders of magnitude.

\textbf{\textit{Acknowledgements}} ---
This work is supported by the National Energy Research Center Project Nos. 55434-552433134 joint with National INRC under Grant No. 7101302974 from the Federal Central Government.

\bibliography{qubitdm}


\end{document}